%% file: main.tex
\documentclass[twocolumn]{aastex63}

\usepackage{color}
\usepackage{CJK}
\usepackage{graphicx}
\usepackage{amsmath}
\usepackage{amssymb}
\usepackage{appendix}

\usepackage{lineno}

\def\ra#1#2#3{#1$^{\rm h}$#2$^{\rm m}$#3$^{\rm s}$}
\def\dec#1#2#3{$#1^\circ#2'#3''$}
\def\nod{\nodata}

\def\pbeta{{\texttt{Prospector}-$\beta$}}

\graphicspath{{./}{figures/}}

\shorttitle{Short GRB Dwarf Host Galaxies}
\shortauthors{Nugent et al.}

\begin{document}

\title{A Population of Short-duration Gamma-ray Bursts with Dwarf Host Galaxies} 

\correspondingauthor{A. E. Nugent}
\email{anyanugent2023@u.northwestern.edu}

\input{affiliation.tex}
\input{authors.tex}

\begin{abstract}
We present a population of 11 of the faintest ($> 25.5$~AB~mag) short gamma-ray burst (GRB) host galaxies. We model their sparse available observations using the stellar population inference code \texttt{Prospector}-$\beta$ and develop a novel implementation to incorporate the galaxy mass-radius relation. Assuming these hosts are randomly drawn from the galaxy population and conditioning this draw on their observed flux and size in few photometric bands, we determine that these hosts have dwarf galaxy stellar masses of $7.0\lesssim\log(M_*/M_\odot)\lesssim9.1$. This is striking as only 14\% of short GRB hosts with previous inferred stellar masses had $M_* \lesssim 10^{9}\,M_{\odot}$. We further show these short GRBs have smaller physical and host-normalized offsets than the rest of the population, suggesting that the majority of their neutron star (NS) merger progenitors were retained within their hosts. The presumably shallow potentials of these hosts translate to small escape velocities of $\sim5.5-80$~km~s$^{-1}$, indicative of either low post-supernova systemic velocities or short inspiral times. While short GRBs with identified dwarf host galaxies now comprise $\approx 14\%$ of the total {\it Swift}-detected population, a number are likely missing in the current population, as larger systemic velocities (observed from Galactic NS population) would result in highly offset short GRBs and less secure host associations. However, the revelation of a population of short GRBs retained in low-mass host galaxies offers a natural explanation for observed $r$-process enrichment via NS mergers in Local Group dwarf galaxies, and has implications for gravitational wave follow-up strategies. 
\end{abstract}

\keywords{short gamma-ray bursts, galaxies, neutron star mergers, dwarf galaxies}

\section{Introduction}
\label{sec:intro}
The astrophysical sites of heavy $r$-process element ($A > 130$) production have implications for the chemical enrichment and evolution of the Universe. Currently, the only observed production sites for $r$-process elements are neutron star (NS) mergers \citep{cbk+17, kmb+17, mhh+17, Pian17, ssd+2017, Tanvir17}. However, while the first NS merger GW170817 was discovered in an old, massive and quiescent host galaxy \citep{bbf+17, pht+17, llt+2017, kfb+2022}, evidence for $r$-process elements has also been discovered in different types of environments, including nearby low-metallicity dwarf galaxies and Galactic metal poor stars \citep{elp+1989, McWilliam1995, Shetrone2001, Venn2012, jfc+2016, cfb+2018, hbp+2018}. The abundances in these latter environments are challenging to explain with NS mergers alone. In particular, the occurrence rates of NS binaries may be too low to create all $r$-process abundances especially at low-metallicity \citep{astq2004, tisa2015} and the expected delay-times for the majority of NS mergers are too long for significant contributions in these young environments \citep{astq2004, dbk+2012, wpt2015, am19, Zevin+DTD}. Rather, supernovae (SNe) and collapsars \citep{mw99}, which can occur on rapid timescales (stellar evolutionary timescales of $\lesssim$~few Myr) with higher occurrence rates in low-metallicity environments have been used to explain the abundances \citep{quian2020, astq2004, tisa2015, sbm2019, shsc2019, Brauer2021}, although there is still no observational evidence that they produce $r$-process elements \citep{bvc+2023}.

Despite support for a faster channel than NS mergers to explain the $r$-process elements in some dwarf galaxies, one such environment may have indeed been enriched from an NS merger event: the $\sim 10^4 M_\odot$, low-metallicity Local-Group dwarf galaxy, Reticulum II. This dwarf galaxy exhibits $r$-process enrichment in several of its brightest stars \citep{jfc+2016, jfsc2016} with yields suggestive of being derived from a single NS merger event, rather than a normal core-collapse supernova (CCSN) \citep{bhp2016, ss2017, oti+2018, sra+2019, tyi2020, csl+2021, jbb2021, mrr+2021}, although theoretical models of collapsars with large $r$-process yields have also been used to explain the abundances \citep{sbm2019}. Additionally, evidence for delayed $r$-process production has been discovered in more massive dwarf galaxies or tidally disrupted dwarf galaxies ($\approx 10^5$-$10^9 M_\odot$; e.g. the LMC, Ursa Minor, Gaia-Sausage-Enceladus, and Wukong), with NS mergers being the most probable cause \citep[e.g.,][]{dkak2018, mht+2021, mrr+2021, Reggiani2021, njc+2022, Limberg2023}. If a NS merger was responsible for $r$-process production in some of these dwarf galaxies, it is natural to search for direct evidence of NS mergers in low-mass galaxies.

Short-duration gamma-ray bursts (GRBs) offer a promising route as the majority are likely spawned from NS mergers \citep{aaloc+17, gvb+17, sfk+2017}, and they are routinely observed over a range of cosmological distances ($0.01 \lesssim z \lesssim 3.0$; \citealt{ber14, skm+18, pfn+20, BRIGHT-I, BRIGHT-II, otd+2022}). However, despite the $\gtrsim 150$ short GRBs detected with NASA's Neil Gehrels \textit{Swift} Observatory ({\it Swift}; \citealt{ggg+04}) and the 84 events with robust host galaxy associations  \citep{vlr+05,ffp+05,bfp+07,dmc+09,fbc+13,ber14,pas19, BRIGHT-I, otd+2022}, there is an apparent lack of galaxies at stellar masses of $\lesssim 10^{8}\,M_{\odot}$, and only $\approx 14\%$ of all {\it Swift} short GRBs are have stellar masses of $\lesssim 10^{9}\,M_{\odot}$ \citep{BRIGHT-II}. Instead, short GRB hosts generally trace the luminosities, star formation rates (SFRs), and metallicities of the typical star-forming field galaxy population, with $\approx 15\%$ in less active galaxies \citep{lb10,BRIGHT-II}. On the other hand, the host galaxies of long-duration GRBs and CCSNe, which originate from massive stars, are comprised of $\approx 35$\% dwarfs \citep{sys+2021, tp2021}. 

Given the strong enrichment of $r$-process elements in Reticulum II, this additionally requires an NS binary to have small NS natal kicks ($\lesssim 15$~km~s$^{1}$; \citealt{bhp2016, bl2016}) to be retained to the dwarf galaxy center and not overcome its relatively small escape velocity. This, however, is contradictory to the larger galactocentric offsets of short GRBs ($\approx 5.6-7.7$~kpc; \citealt{cld+11, fb13, tlt+14, BRIGHT-I, otd+2022}), which likely have progenitors with larger natal kicks (c.f., \citealt{zkn+2019,pb2021}). Furthermore, inferences on the delays of star formation episodes in dwarf galaxies \citep{bsw2015, jfb2015, jsr+2023} suggest that the $r$-process producing event in Reticulum II likely has a delay time of $\lesssim 100$~Myr, at odds with the inferred minimum delay times from host stellar populations of $\approx 200$~Myr \citep{Nakar2006, bfp+07, Jeong2010, Hao2013, Wanderman2015, Anand2018, Zevin+DTD} and the observed population of Galactic binary NS (BNS) systems \citep{tkf+17,am19}. However, predictions made from stellar population synthesis and models of the delay time distribution (DTD) of Galactic BNS systems estimate that the minimum delay time can be as low as $\sim 10$~Myr, \citep{bkv2002, dbk+2012, vns+18, bp2019}. 

Current short GRB host samples are generally limited to the galaxies with luminosities of $\gtrsim 10^{9} L_{\odot}$ especially beyond $z \gtrsim 1$ (e.g., \citealt{BRIGHT-I, otd+2022}) and it has been challenging to overcome the bias against identifying high-redshift ($z \gtrsim 1.5$) and/or low-luminosity hosts. To fill this gap and explore a possible missing dwarf host population, here we present modeling of 11 faint short GRB hosts ($\gtrsim 25.5$~mag) that have been absent in previous stellar population modeling studies, to estimate their redshifts and stellar masses. By default, these hosts have limited observational data, requiring novel stellar population modeling techniques in order to put useful constraints on their properties. This sample represents all remaining short GRBs with robust host associations that do not have previous stellar population modeling results, but for which it is possible with novel stellar population modeling techniques. We discuss our host sample and the available observations in Section~\ref{sec:sample}. In Sections~\ref{sec:sp_models}-\ref{sec:sp_results}, we detail our stellar population modeling and results. In Section~\ref{sec:grb_prop}, we examine any trends with respect to short GRB properties, including the $\gamma$-ray properties, afterglow luminosities, and galactocentric offsets. We discuss selection effects, delay times, and implications for this population and GW follow-up in Section \ref{sec:disc}. Finally, we summarize our main conclusions in Section \ref{sec:conclusion}. 

Unless otherwise stated, all observations are reported in the AB magnitude system and have been corrected for Galactic extinction in the direction of the GRB \citep{MilkyWay,sf11}.  We employ a standard WMAP9 cosmology of $H_{0}$ = 69.6~km~s$^{-1}$~Mpc$^{-1}$, $\Omega_\textrm{M}$ = 0.286, $\Omega_\textrm{vac}$ = 0.714 \citep{Hinshaw2013, blw+14}. 

\section{Host Sample and Observations}
\label{sec:sample}
We start with the host sample described in \citet{BRIGHT-I}, which includes host galaxy associations for 84 short GRBs discovered by \textit{Swift} with afterglows detected to  $\lesssim 5\arcsec$ localization and not along high-Galactic extinction sitelines ($A_V < 2$~mag) that would impair possible host detection. While 69 hosts had sufficient data to be modeled in \citet{BRIGHT-II}, there are 14 remaining that do not have determined stellar population properties. Of these, 11 comprise the faintest detected hosts in the entire sample, with optical and near-IR magnitudes $\gtrsim 25.5$~AB~mag (only four other hosts from the parent sample of 69 have comparable optical magnitudes, but with determined photometric redshifts). These 11 hosts, listed in Table \ref{tab:obs}, are almost exclusively detected with NASA's \textit{Hubble Space Telescope} (\textit{HST}) in only one or two photometric filters and constitute the host sample in this work. Of the three hosts not included in our sample, one is a low-redshift GRB in a crowded field (GRB\,080905A) and the other two (GRBs 081226A and 160601A) do not have \textit{HST} observations for effective radii measurements, which are used in this analysis (see Section \ref{sec:sp_modif}). 

According to \citet{BRIGHT-I}, the majority of hosts in this faint sample are classified as ``Gold" host associations (probabilities of chance coincidence, $P_{cc} < 0.02$), with three hosts as ``Silver" (GRBs 080503, 131004A, and 150424A; $0.02 < P_{cc} < 0.09$), and two as ``Bronze'' (GRBs 091109B and 130912A; $0.09 < P_{cc} < 0.20$). We list the association type in Table~\ref{tab:obs}. Two of the hosts, furthermore, have spectroscopic redshifts determined from their GRB afterglows: GRB\,090426A ($z=2.609$) and GRB\,131004A ($z=0.717$), but undetermined stellar masses. We also include two new photometric detections for the host of GRB\,211106A \citep{fbd+2023}. For the remaining nine host galaxies, we determine upper limits on the redshifts from the optical afterglow detections (available for all GRBs except GRB\,211106A), corresponding to a lack of suppression blueward of the Lyman-$\alpha$ limit \citep{fbm+15}. This effectively places upper limits on their redshifts of $z \approx 1-4$ depending on the burst.

\begin{figure}[t]
\centering
\includegraphics[width=0.49\textwidth]{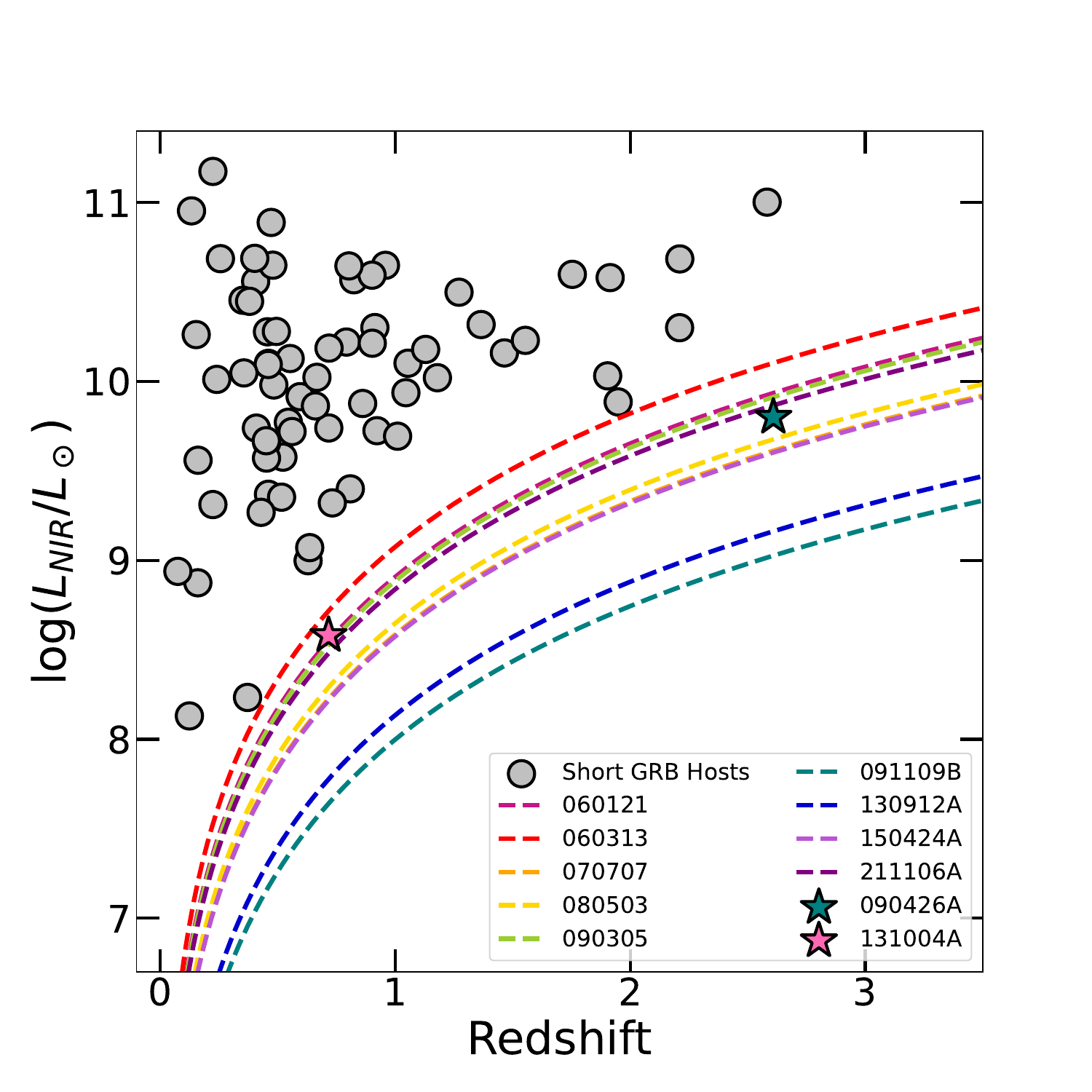}
\vspace{-0.3in}
\caption{The observed or inferred $J$-band luminosities and redshifts of the sample of short GRB host galaxies studied in \citet{BRIGHT-II} (grey circles). We plot the observed luminosities for the faint short GRB host sample in Table \ref{tab:obs}, using either the F110W filter or the filter closest to the $J$-band central wavelength for each short GRB. Dashed lines represent the nine hosts with no known redshifts and stars represent the hosts with known redshifts. We see that the faint hosts sample are generally less luminous across all redshift than the rest of the host sample, implying they likely represent a unique group of low-luminosity, low-mass environments.}
\label{fig:j_lum}
\end{figure}

In Figure~\ref{fig:j_lum}, we show the near-IR luminosities of the faint host galaxies as a function of redshift, compared to the population with previously determined stellar population properties \citep{BRIGHT-II}. For the latter sample, in the absence of relevant near-IR data, we use the model SEDs derived in \citet{BRIGHT-II} to derive a $J$-band magnitude.  We note that these faint hosts represent the lowest luminosity host galaxies to date and stand in contrast to the rest of the host population, out to a maximum redshift $z \approx 4$, where they begin to appear more similar to the current host sample ($\log(L_\textrm{NIR}/L_\odot) \approx 10.1$). As near-IR luminosities are strongly correlated to stellar mass, this hints that unless they are all at high-redshift, they likely have smaller stellar masses as well.

We collect \textit{HST} and \textit{VLT} photometry and upper limits from NASA's \textit{Spitzer} Space Telescope in \citet{fbf10}, \citet{fb13}, \citet{BRIGHT-I}, and \citet{fbd+2023}. For GRBs\,060121, 060313, and 080503 (only F160W filter), we perform aperture photometry using standard tasks in IRAF/{\tt phot}, whereas such measurements are available for the other hosts in our sample.  We further collect the effective radii ($r_e$) from \citet{BRIGHT-I} which are available for all of the host galaxies in our sample. All photometry, spectroscopic redshifts (when available), maximum possible redshifts, and $r_e$ measurements are listed in Table~\ref{tab:obs}.  

\input{grb_obs}

\section{Stellar Population Modeling}
\label{sec:sp_models}

\subsection{Prospector-$\beta$}
\label{sec:pbeta_models}
To model the stellar population properties of the host galaxies, we use the Python-based spectral energy distribution (SED) code \pbeta\ \citep{prospector_beta} over all photometric detections and upper limits of the hosts. \pbeta\ was specifically designed to infer the redshifts and stellar masses of faint galaxies with limited photometric wavelength coverage, and thus far has been used on faint \textit{HST} and \textit{James Webb Space Telescope} (JWST) targets \citep{wll+2023}. In contrast to other versions of \texttt{Prospector} \citep{Leja2019, jlc+2021}, it uses non-uniform priors on key stellar population properties, including mass formed ($M_F$), redshift ($z$), stellar metallicity ($Z_*$), and the star formation history (SFH), informed from mock catalogs and observations \citep{williams2018, Leja2020}, which greatly reduces the parameter space to only stellar population property solutions that are consistent with observed galaxy populations in deep extragalactic fields. Indeed, it has already been shown that the \pbeta\ non-uniform priors better estimate the stellar masses of mock galaxies with limited photometric data than uniform priors \citep{prospector_beta}. With limited observational data (the majority with $\leq 2$ data points each), our sample of 11 faint short GRB hosts represents an excellent test data set for \pbeta.  However, while \pbeta\ can give robust estimates on redshift and stellar mass with very few photometric detections, it is not expected to constrain $Z_*$ and the SFH, as these are best determined through modeling the shape of an SED across a wide range of wavelengths. These are nonetheless included in the fits, not in the hopes of producing useful constraints on them, but rather in order to properly marginalize over them. \pbeta\ fits the observed photometry of a galaxy to model SEDs produced through \texttt{FSPS} and \texttt{python-fsps} \citep{FSPS_2009, FSPS_2010}, which, by default, uses \texttt{MIST} models \citep{MIST} and \texttt{MILES} spectral library \citep{MILES}. We apply the nested-sampling fitting routine \texttt{dynesty} to derive posterior distributions of the stellar population properties of interest, including $M_F$, $z$ (when it is not already known), $Z_*$, and the SFH. 

\begin{figure*}[t]
\centering
\includegraphics[width=0.8\textwidth]{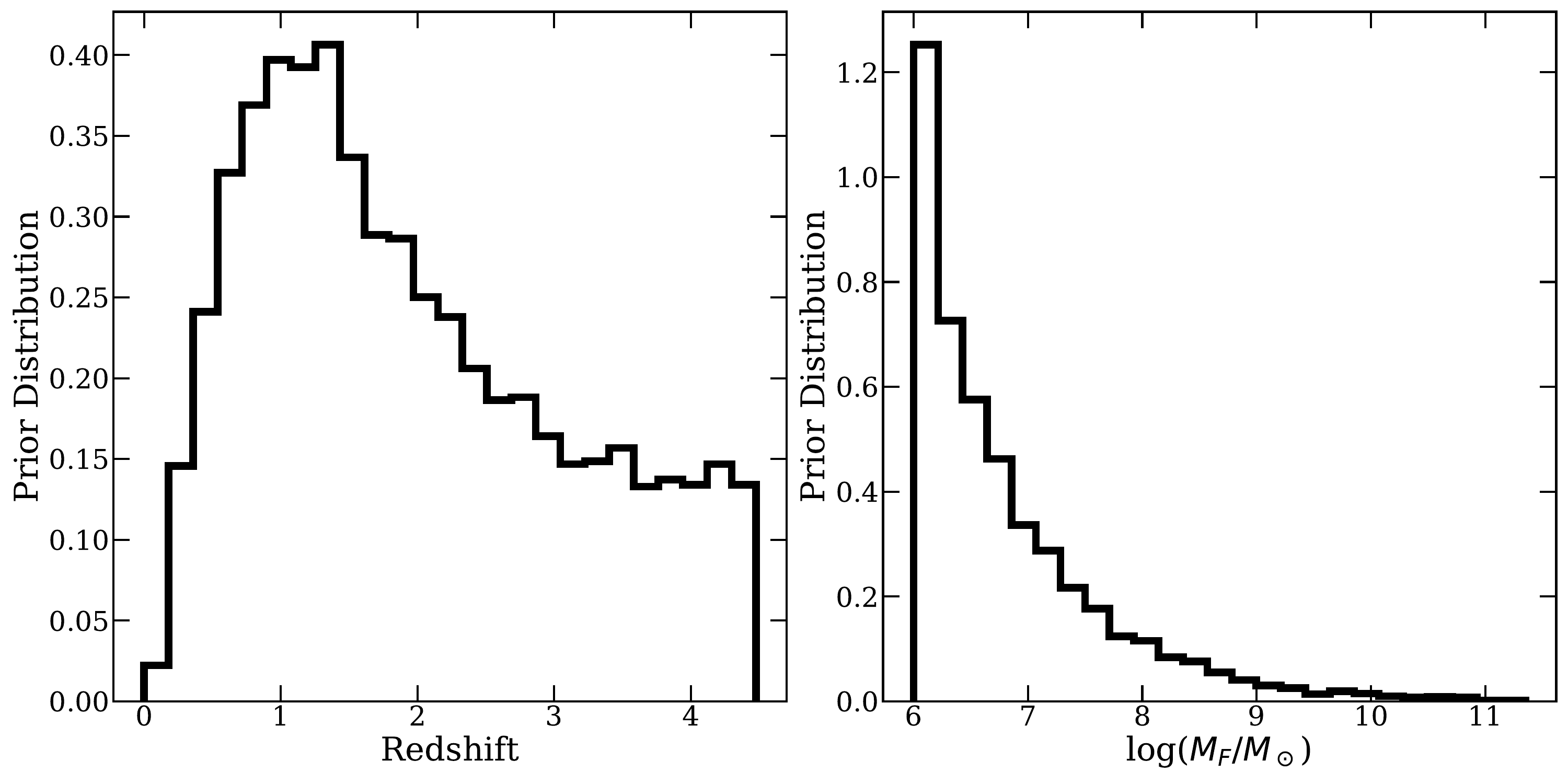}
\caption{The \texttt{Prospector}-$\beta$ priors on redshift ($z$) and mass formed ($M_F$) \citep{prospector_beta}. The non-uniform priors better predict stellar population properties of faint galaxy targets with limited photometric coverage, as they limit the sampling to only plausible solutions based on observed galaxy trends.}
\label{fig:pbeta_prior}
\end{figure*}

\pbeta\ contains several different model templates, which have various combinations of prior distributions. For the faint hosts with no known redshifts, we use the \texttt{NzSFH} model template. The \texttt{NzSFH} template employs a redshift prior that is based on the number density of galaxies across $0 \leq z \leq 15$, given by Equation 2 in \citet{prospector_beta}, and shown in Figure \ref{fig:pbeta_prior}. We modify the redshift prior by placing a maximum given by the detection of the optical afterglow for each short GRB (Table \ref{tab:obs}). We set the maximum redshift for GRB 211106A to $z=4.5$ as this GRB has no detected optical afterglow and we do not expect short GRBs to be detected by \textit{Swift} much greater than this redshift \citep{lsb+16}. The $M_F$ prior in the \texttt{NzSFH} template is dependent on the mass function derived in \citet{Leja2020} at a given redshift $z$ and has a range $10^6 \leq M_F \leq 10^{12} M_\odot$ (shown in Figure \ref{fig:pbeta_prior}). We note that the mass function in \citet{Leja2020} is only constrained down to $\approx 10^8 M_\odot$, depending on the redshift. Thus, the function is extrapolated down to lower stellar masses. As there are relatively more low-mass galaxies in the Universe than higher-mass galaxies, the prior tends to favor lower mass solutions, although it does not forbid exotic parameter spaces given sufficiently convincing data; for example, it has been used to make recent discoveries from {\it JWST} of high-mass galaxies at higher redshifts \citep{labbe2023}. We note that $M_F$ is converted to a stellar mass $M_*$ within the \texttt{Prospector} infrastructure, which we report hereafter. The \texttt{NzSFH} template furthermore includes a Gaussian prior on $Z_*$, which is dependent on the mass-metallicity relation described in \citet{gcb+05}. It also incorporates a non-parametric SFH that is a function of the age of the Universe at redshift $z$ and the mass formed in the galaxy. We fit for the star formation rate in seven log-spaced age bins to determine the SFH; we refer the reader to a thorough description of the dynamic non-parametric SFH prior and an explanation for how the age bins are made in \citet{prospector_beta}. Finally, we apply the \citet{KriekandConroy13} dust attenuation model which measures the offset from the \citet{calzetti2000} attenuation curve and the ratio of dust attenuated from old to young stellar light, as this is the standard dust model in \pbeta. For the two faint short GRB hosts that have known redshifts, we employ the \texttt{PhiSFHzfixed} model templates, which contain prior distributions in $M_F$, $Z_*$, and the SFH that are identical to those in the \texttt{NzSFH} template, but allows redshift to be a fixed parameter.

\subsection{Implementation of Mass-Radius Relation}
\label{sec:sp_modif}
By default, we have limited available data for our sample of faint short GRB hosts, making them excellent cases for modeling with \pbeta. Notably, stellar mass and redshift can be robustly constrained in some cases even in the absence of extensive photometric coverage \citep{aca+2023}, if the targets do not have unusual colors or magnitudes given their properties and redshifts. However, given the lack of observational data for this sample of hosts, we wish to expand upon the current \pbeta\ infrastructure to find more informed constraints on the true stellar masses and redshifts using the well-known and characterized galaxy mass-radius relation \citep{vdw2014}. By modifying \pbeta\ to be be informed by the galaxy size, in addition to the integrated flux, we can better constrain the stellar mass estimates to more physical values. We note that while this is a novel implementation in \pbeta, this concept has been tested previously and successfully in \citet{dan+2022}. 

We employ the \citet{vdw2014} mass-radius relation in our analysis. This relation specifically constrains the stellar mass of galaxy given its physical effective radius (radius from within which half of the galaxy light is contained; $r_e$), or vice-versa, at a given redshift and is based on galaxies observed in the CANDELS/3D-HST \citep{candels, Brammer2012} fields with \textit{HST/WFC3} across $0 < z < 3$. Stellar masses for these galaxies were determined through the stellar population synthesis code \texttt{FAST} \citep{kvl+2009} and the $r_e$ were calculated using \texttt{GALFIT} \citep{galfit} over the available \textit{HST/WFC3} data (F814W, F125W, F140W, and F160W filters; \citealt{vdw2012})\footnote{We note that \citet{vdw2014} mass-radius relation was observed for galaxies down to $M_*\approx10^9 M_\odot$. However, it has been shown that this function can be reasonably extrapolated down to lower stellar masses \citep{nhm+2021}, especially for late-type galaxies.}. Although the stellar masses derived from \texttt{FAST} are known to be $\approx 0.1-0.2$~dex smaller than stellar masses inferred from \texttt{Prospector} \citep{Leja2019}, the scatter on the mass-radius relation, which we take into account in our modeling, is the dominant source of uncertainty and outweighs this small systematic offset. 

We describe our novel implementation of the \citet{vdw2014} mass-radius relation in \pbeta\ in Appendix \ref{app:MR}. In essence, we use the observed size (see Table \ref{tab:obs}; \citealt{fbf10, fb13, BRIGHT-I}) and photometry of the host to constrain the stellar population properties, as opposed to just the photometry (Section \ref{sec:pbeta_models}). We do so by determining the likelihood of a host $r_e$ within a distribution of possible galaxy sizes derived at a sampled \pbeta\ stellar mass and redshift in the nested sampling routine. Hereafter, we call the \pbeta\ and mass-radius relation method: \pbeta\ ($M_*-r_e$). 

The inclusion of the \citet{vdw2014} mass-radius relation effectively increases the probability of \pbeta\ samples where the observed size of the galaxy is well-constrained within a distribution of physical sizes from $M_*$ and $z$. These probabilities are maximized when the observed galaxy size is closer to the mean of physical size distribution, rather than on the outskirts. As the uncertainty on the \citet{vdw2014} mass-radius relation increases with redshift, solutions at lower redshifts that are in good agreement with the observed size are likely to be maximized more so than solutions at high redshift, where the observed size is further from the mean of the distribution derived from the sampled $M_*$ and $z$.
This approach implicitly assumes that GRB hosts are typical members of the galaxy population and if they instead only occur in highly unusual systems, the constraints will be more difficult to interpret. However, given that we have already seen that short GRB hosts do have similar luminosities, SFRs, stellar masses, and metallicities to field galaxy populations \citep{BRIGHT-I, BRIGHT-II, otd+2022}, we find that this underlying assumption is likely legitimate. 

\input{results}

\section{Host Galaxy Properties}
\label{sec:sp_results}

\input{resampled}

Here we present and compare the results of the stellar population fitting from both \pbeta\ and \pbeta\ ($M_*$-$r_e$); we list the redshift and stellar mass medians and 68\% confidence intervals from these fits in Table \ref{tab:results} and show the posterior distributions of the 9 hosts without previously known redshifts in Figure \ref{fig:resampled}. First, when comparing the \pbeta\ determined posterior distributions in stellar mass and redshift to the prior distributions (see Figure \ref{fig:pbeta_prior}), we find that with the exception of GRB\,060121,  all posterior distributions are distinct from the prior distributions, suggesting that the data are providing meaningful constraints and the posteriors are not simply tracing the prior. Since GRB\,060121 is one of the faintest in the sample, it is somewhat unsurprising that the data may not be sufficient to provide strong deviations from the prior. 

For the \pbeta-modeled population, we find a redshift median and 68\% confidence interval on the population of $z = 0.73^{+1.35}_{-0.32}$ and stellar mass median and 68\% confidence interval of log($M_*/M_\odot$)=$7.99^{+1.0}_{-0.85}$. While the redshifts are comparable to those found for the short GRB population \citep{BRIGHT-I,BRIGHT-II}, the stellar masses for the faint host population are lower than the rest of the observed host population, which have a median and 68\% confidence interval of log(M$_*$/M$_\odot$)=$9.69^{+0.75}_{-0.65}$ \citep{BRIGHT-II}. 

\begin{figure*}[t]
\centering
\includegraphics[width=0.49\textwidth]{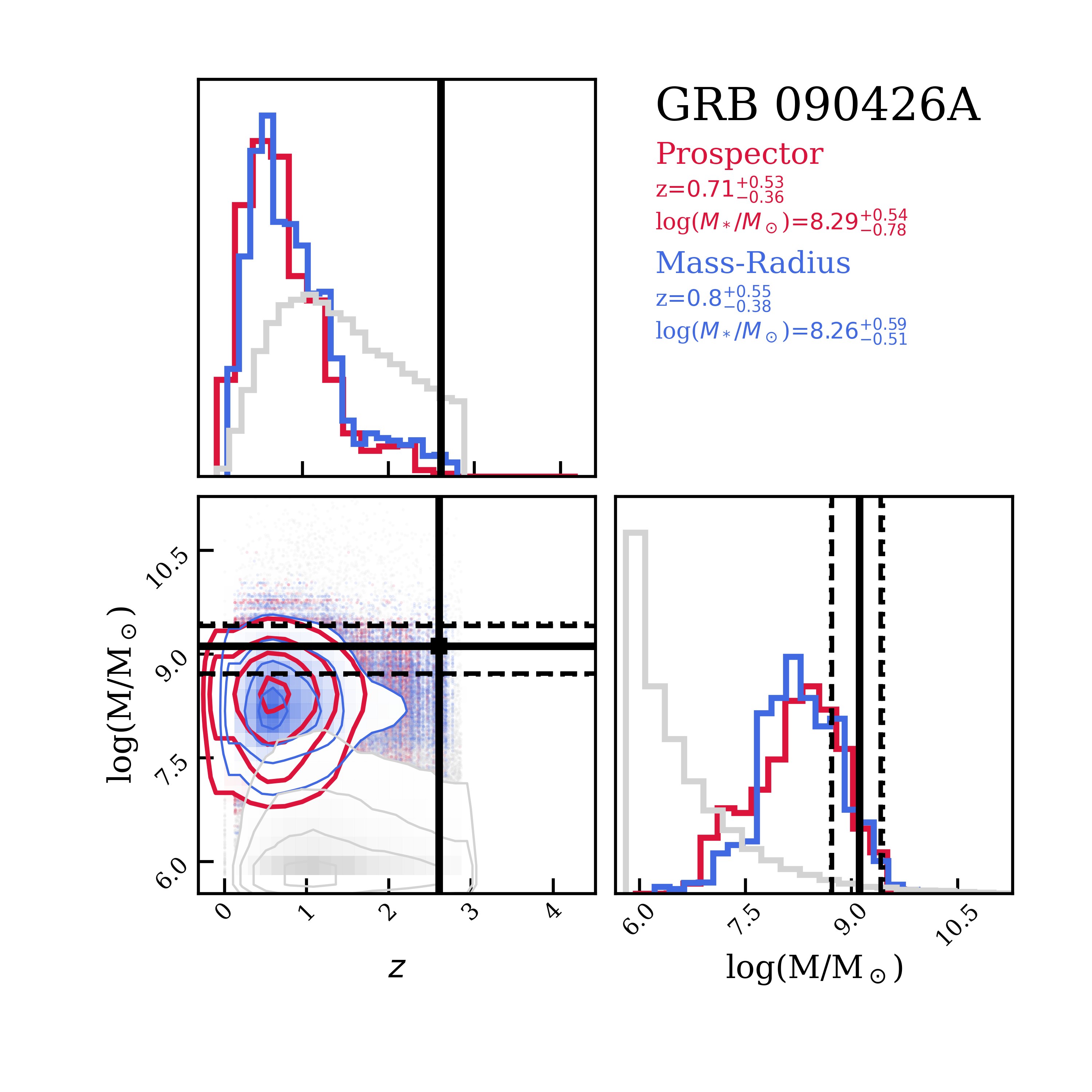}
\includegraphics[width=0.49\textwidth]{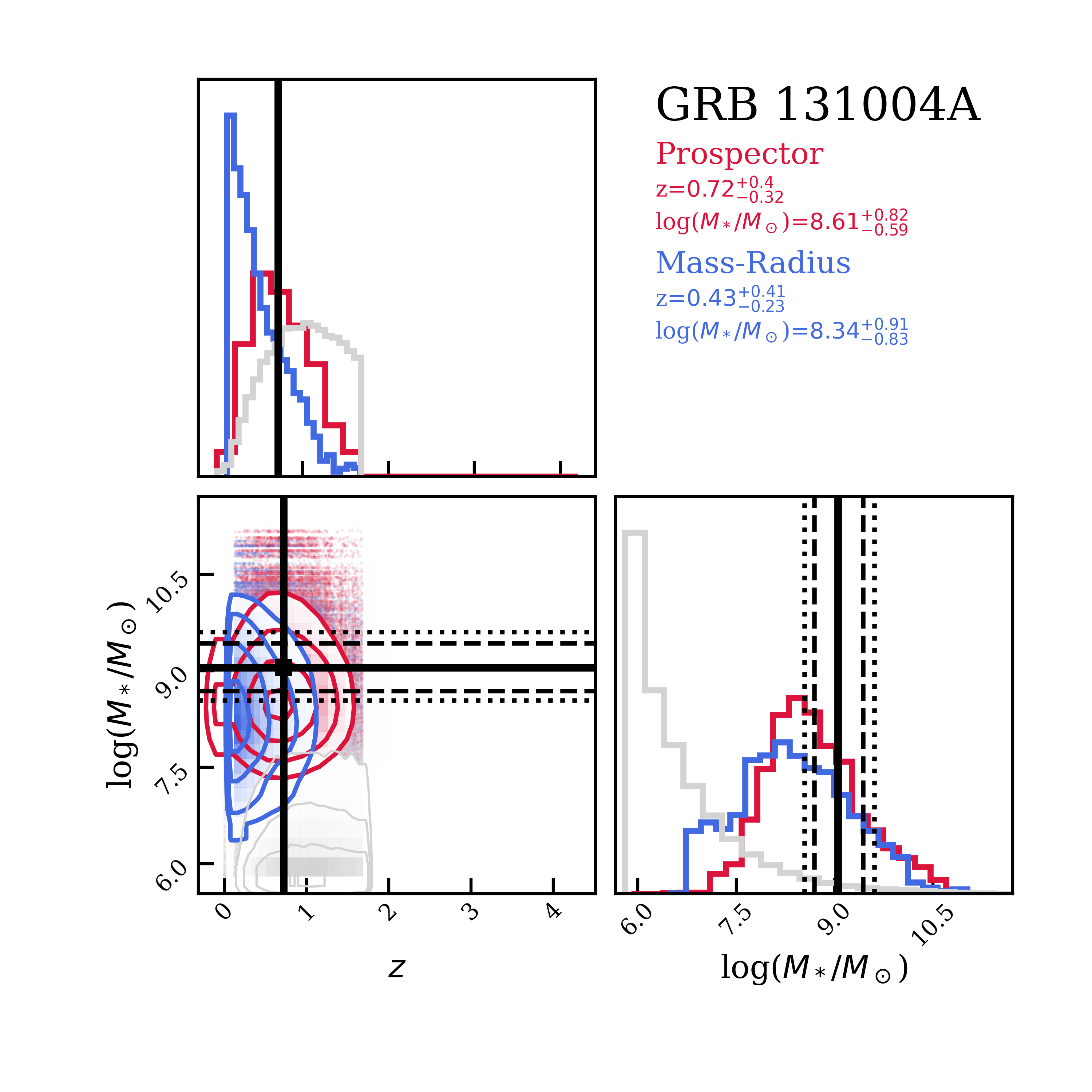}
\vspace{-0.1in}
\caption{The stellar mass and redshift estimates results for the \pbeta\ (red) and \pbeta\ ($M_*$-$r_e$) (blue) test cases for the two faint hosts with known redshifts: GRBs 090426A (\textit{left}) and 131004A (\textit{right}). The prior distributions are shown in grey. We plot the median of their stellar mass estimates from the fits using their known redshifts (black line), along with the 68\% confidence region for the \pbeta\ fit (dashed black lines) and the \pbeta\ ($M_*$-$r_e$) fit (dotted black lines). We find that while the stellar masses determined in the test cases are consistent with the fits using the known redshifts, the true redshift was harder to constrain in the test cases.}
\label{fig:test_cases}
\end{figure*}

For the \pbeta\ ($M_*$-$r_e$) fits, we find that slightly lower redshifts are preferred (median and 68\% confidence interval decreases to $z = 0.54^{+0.47}_{-0.29}$). The stellar masses remains fairly consistent with the original fits, with a median and 68\% confidence interval of log($M_*/M_\odot$)=$7.82^{+1.2}_{-0.86}$, although still lower than the rest of the short GRB population. Comparing the results of the \pbeta\ and \pbeta\ ($M_*$-$r_e$) fits (Figure \ref{fig:resampled}), we note that while the majority of results shift towards a lower redshift solutions, the stellar masses stay consistent within the 68\% confidence intervals. Given that the majority of this population falls at $M_* \lesssim 10^9 M_\odot$, the stellar mass limit of dwarf galaxies \citep{bb2017}, these results strongly suggest that this is a population of dwarf galaxies. Indeed, $\approx 84\%$ of the \pbeta\ ($M_*$-$r_e$) posterior distributions across all 11 hosts result in $M_* \lesssim 10^9 M_\odot$. We further note that even when we only analyze the population of very robust host associations (Gold Sample; Table \ref{tab:obs}), the stellar mass stays similarly low, at log($M_*/M_\odot$)=$7.76^{+1.27}_{-0.78}$, hinting that NS mergers are indeed occurring in low mass environments.

\subsection{Testing the \pbeta\ Framework}
\label{sec:test}
To better understand the stellar mass and redshift results for our faint short GRB hosts, we perform \pbeta\ fits and \pbeta\ ($M_*$-$r_e$) setting the redshift free for the two GRBs with known redshifts, GRBs\,090426A and 131004A, and compare these results to the fixed redshift results (which represents the likely ``true'' results). In Figure~\ref{fig:test_cases}, we show the posterior distributions of their redshifts and stellar masses for the \pbeta\ fits and \pbeta\ ($M_*$-$r_e$) fits and overplot the ``true'' results as black lines. We find that the true redshift for GRB 090426A ($z=2.609$) lies near the 99th percentile of, but still within, the evolving galaxy population after conditioning on the observed flux and size, while the stellar mass is fairly consistent. This is an interesting result and may suggest that the host of this GRB has other unusual properties compared to the normal galaxy population. Though possibly counter-intuitive, we also note that it is possible for the inferred stellar mass to stay consistent across a range of redshifts, as at higher redshifts, galaxy colors tend to be bluer, which decreases their mass-to-luminosity ratio, while their intrinsic luminosity is higher due to cosmological dimming. Subsequently, stellar mass estimates can stay flat over a range of redshifts. In the case of GRB131004A, the true redshift ($z = 0.717$) and stellar mass estimate falls within the 68\% confidence interval for both fits, suggesting this may be a more typical galaxy in terms of its photometric properties and size. Taken together, these test cases show that stellar mass is a fairly robust parameter, but redshift may be more challenging to constrain especially when a galaxy's properties deviate from the normal population. Given this insight, we emphasize that the dwarf host implication for the sample studied here is credible.

\begin{figure*}[t]
\centering
\includegraphics[width=0.49\textwidth]{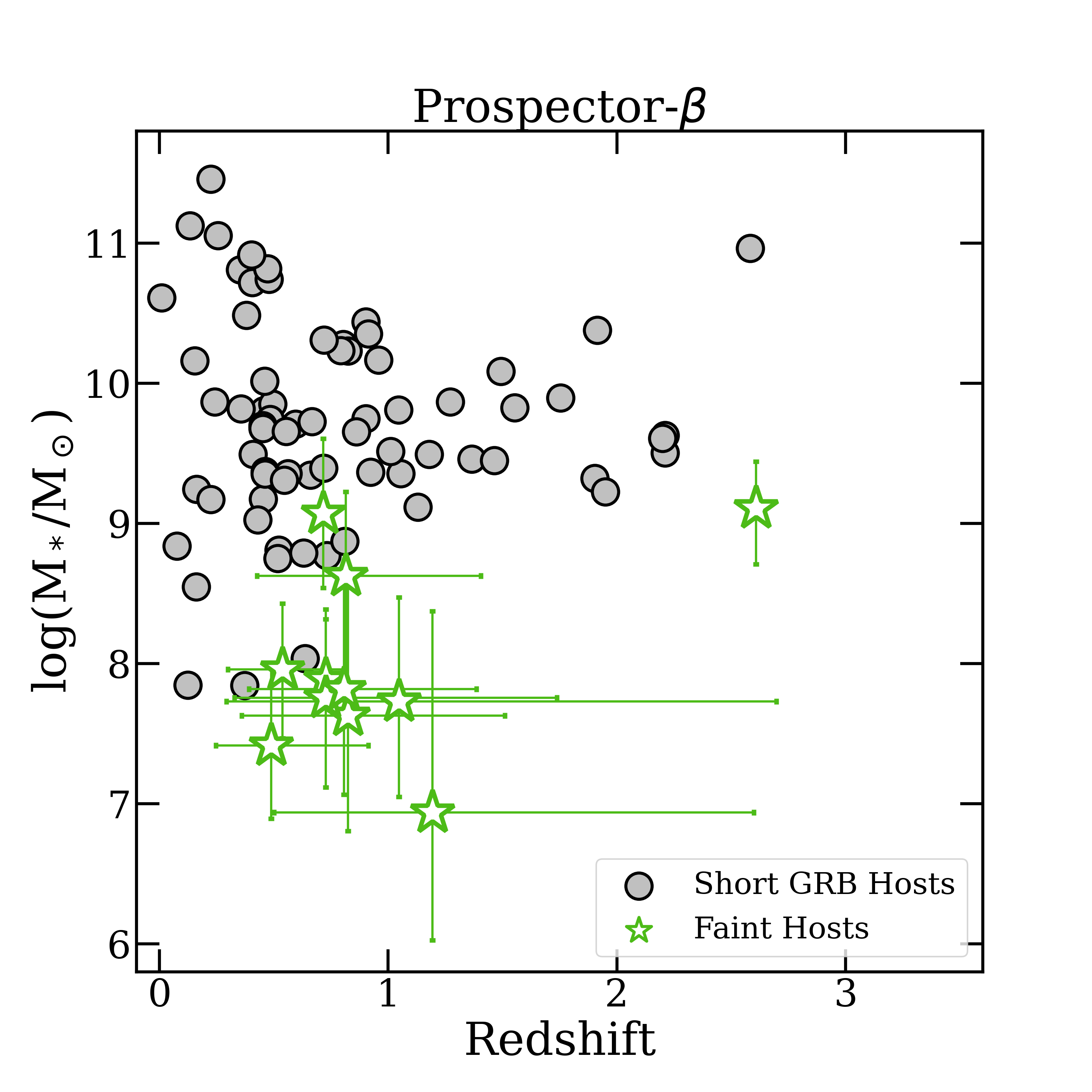}
\includegraphics[width=0.49\textwidth]{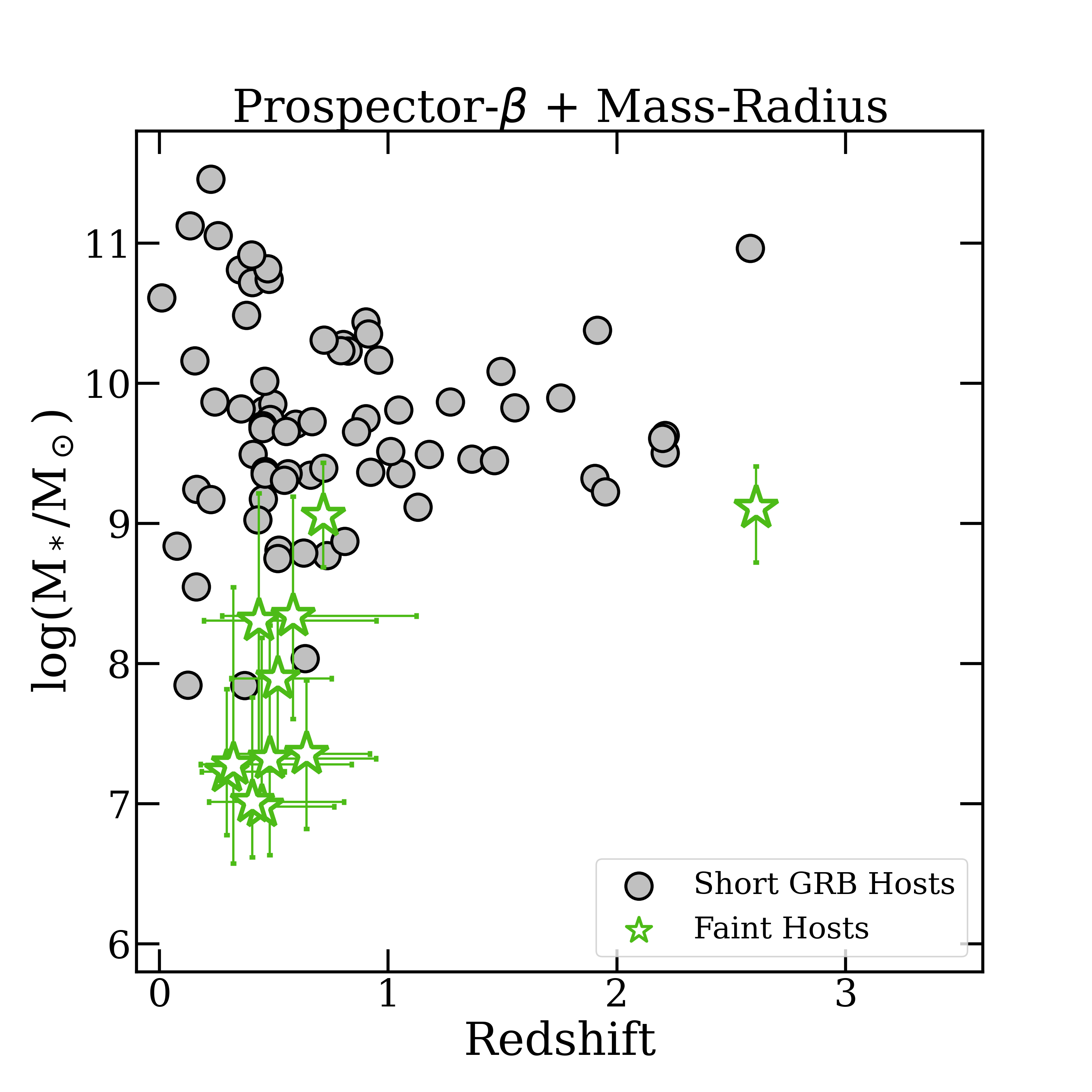}
\vspace{-0.1in}
\caption{\textit{Left:} The stellar masses (log($M_*/M_\odot)$) and redshifts of the faint host galaxy population (green stars) estimated from the \pbeta\ fits, in comparison to the full short GRB host sample, with the same coloring as Figure \ref{fig:j_lum}. \textit{Right:} The same figure, but with stellar masses and redshifts estimated from the \pbeta\ ($M_*$-$r_e$) fit. We see in both cases this population trends towards lower stellar masses than the full host sample, with more moderate redshifts in-line with expectations from the full host sample.}
\label{fig:mass_redshift}
\end{figure*}

\subsection{Comparison to the Full Short GRB Host Sample}
\begin{figure}[t]
\centering
\includegraphics[width=0.49\textwidth]{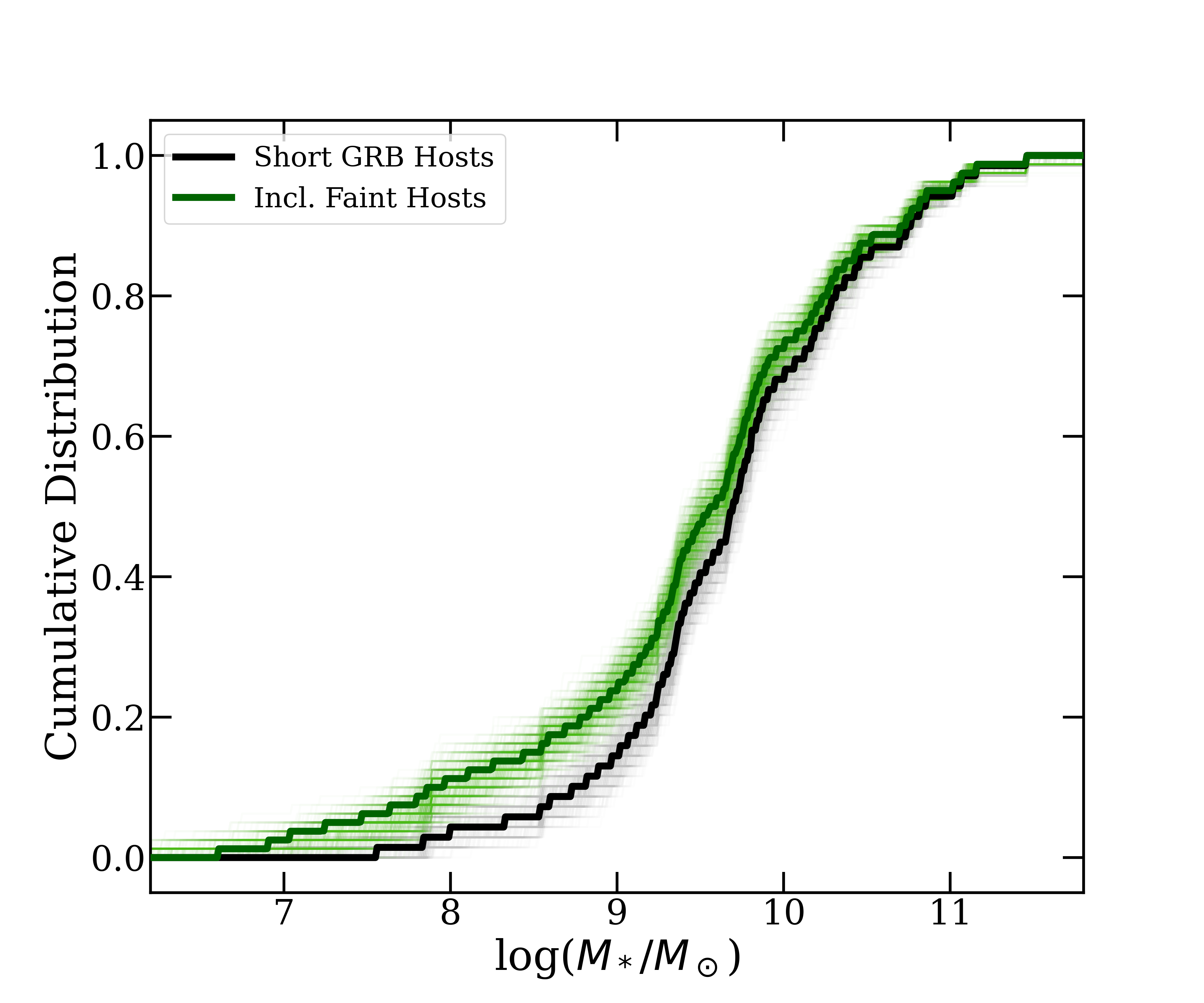}
\vspace{-0.1in}
\caption{The cumulative distribution (CDF) in stellar of the short GRB sample in \citet{BRIGHT-II} (grey lines) and the affect of including the faint host short GRB sample (green lines). We only show the affect from the \pbeta\ ($M_*$-$r_e$) fits as both stellar population modeling methods result in the same affect to the stellar mass distribution. The darker lines represent the median of the CDF and the lighter lines show 5000 realization on the CDF. We find that while the stellar mass distribution stays similar to the result in \citet{BRIGHT-II}, the tail of the CDF is extended toward lower masses.}
\label{fig:new_mass_cdf}
\end{figure}

We compare the redshift and stellar masses of the faint host population to those of the entire short GRB host population in Figure \ref{fig:mass_redshift}. Strikingly, while the large majority of short GRB hosts have stellar masses of $M_*\gtrsim 10^{9}\,M_{\odot}$, our entire sample falls near or below this value, regardless of the inclusion of the $M_*$-$r_e$ relation; indeed, only 14\% of the 69 hosts with stellar mass estimates in \citet{BRIGHT-II} have $\log(M_*/M_\odot)\lesssim 9$. Including our faint host population studied here in the stellar mass distribution of all short GRB hosts (to make a total of 80 modeled) does not significantly shift the population median, but extends the low-mass tail, as shown in Figure \ref{fig:new_mass_cdf}. We find that both the \pbeta\ and \pbeta\ ($M_*$-$r_e$) fits change the stellar mass median and 68\% confidence interval of the entire short GRB host population to log(M$_*$/M$_\odot$)=$9.57^{+0.78}_{-1.02}$. Including the 10 hosts in \citet{BRIGHT-II} that can be classified as dwarf galaxies ($M_*\lesssim 10^9 M_\odot$; \citealt{bb2017}), we find that out of the population of $\sim 155$ \textit{Swift} short GRBs detected over 2005-2023 \citep{lsb+16}, $\approx 13.5\%$ occur in dwarf galaxies (with $\approx 8\%$ in galaxies with $M_* \lesssim 10^8 M_\odot$). We are likely missing some short GRBs in dwarf galaxies (see Section~\ref{sec:disc_select}), so these percentages probably represent lower limits. In fact, the true fraction could in reality be more comparable to the population of field galaxies that have $M_* \lesssim 10^9 M_\odot$, which is $\approx 20\%$ of total galaxy population estimated from the Local Volume Legacy Survey \citep{lgk+2011, tp2021}.

Overall, our results show that these events very likely occurred in low-mass environments, and that the host galaxy population of short GRBs is more diverse than previously thought.

\section{Short GRB Properties}
\label{sec:grb_prop}
\subsection{Optical Afterglow Luminosities and Offsets}
\label{sec:AG_offsets}
We next explore properties of the faint host sample to determine if there are any other distinguishing features that set them apart from the rest of the short GRB host population. We first compare their optical afterglow luminosities and galactocentric offsets. As afterglow luminosity generally scales with circumburst density \citep{gs02}, it can be used as a proxy for the burst environment. It was also found that short GRBs at larger offsets generally have fainter observed afterglows, again a likely byproduct of decreasing ISM density at larger offsets \citep{pb02,ber10}. Thus, by exploring a combination of luminosity and offsets, we can probe the local properties of the faint host sample.

First, all short GRBs in the faint host sample have optical afterglow detections, except for GRB\,211106A\footnote{This event had both bright millimeter and radio detections, and a possible explanation for its optical darkness is high dust extinction surrounding the event \citep{laskar_2022, fbd+2023}.}. This is a much higher fraction of events with an optical afterglow detection than the full short GRB population (only $\approx 30\%$; \citealt{ber14,fbm+15}). However, this is likely a product of how this sample was selected, as it becomes increasingly challenging to make host assignments for extremely faint hosts in the absence of a sub-arcsecond localization (e.g., \citealt{eb2017}).

\begin{figure}[t]
\centering
\includegraphics[width=0.49\textwidth]{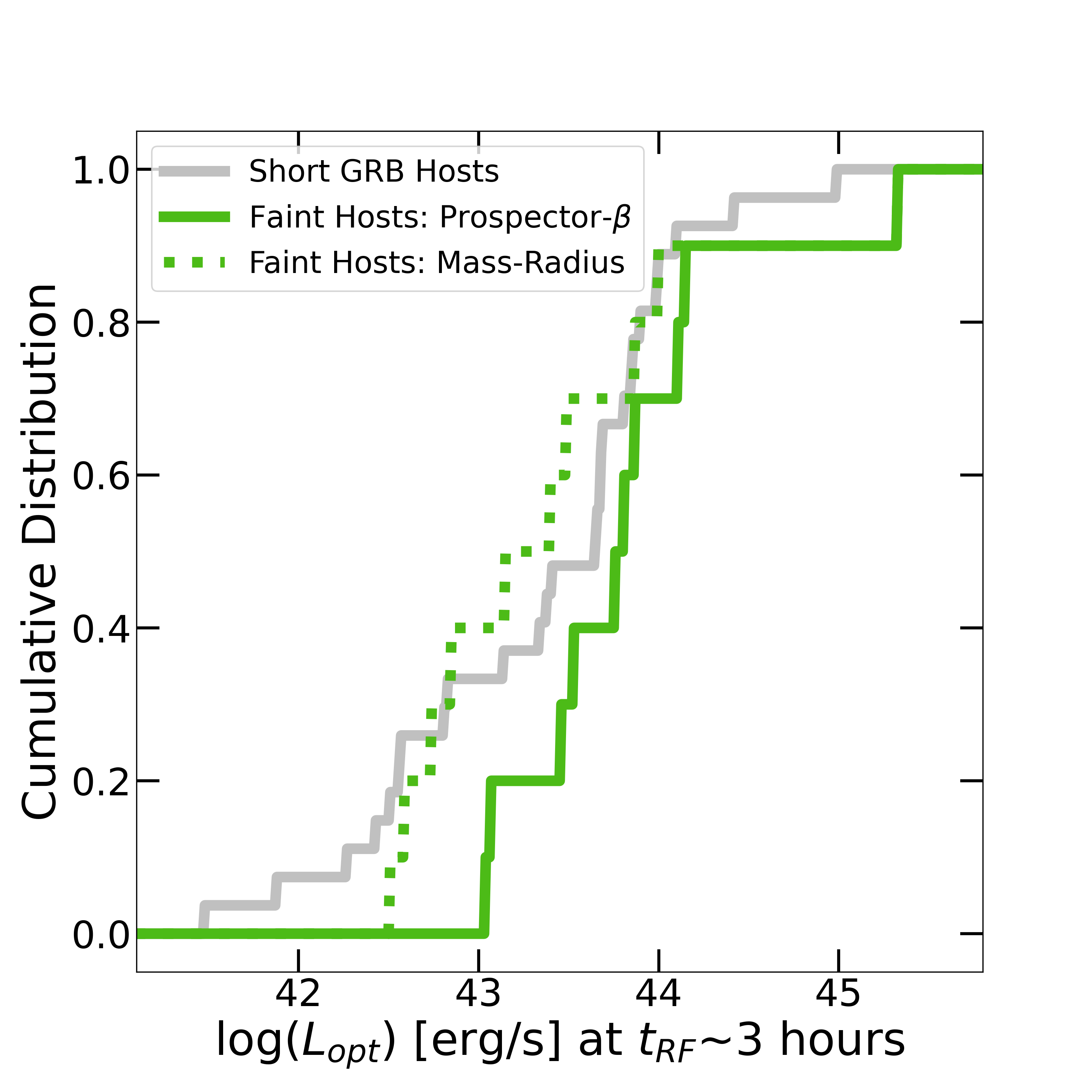}
\vspace{-0.1in}
\caption{The CDFs of the detected optical afterglow luminosities from the faint host short GRBs (green) in comparison the rest of the short GRB population (grey). The straight line represents afterglow luminosities inferred from the \pbeta\ redshifts, and the dotted line represents those inferred from the \pbeta\ ($M_*$-$r_e$) redshifts. The grey line in this case can be treated as an upper limit on the distribution of short GRB optical afterglows, given that it neglects the $\sim60\%$ of short GRBs that only have upper limits. The short GRBs in the faint hosts have similar optical afterglow luminosities than the rest of the short GRBs from their inferred redshifts.}
\label{fig:cdf_AG}
\end{figure}

\begin{figure*}[t]
\centering
\includegraphics[width=0.49\textwidth]{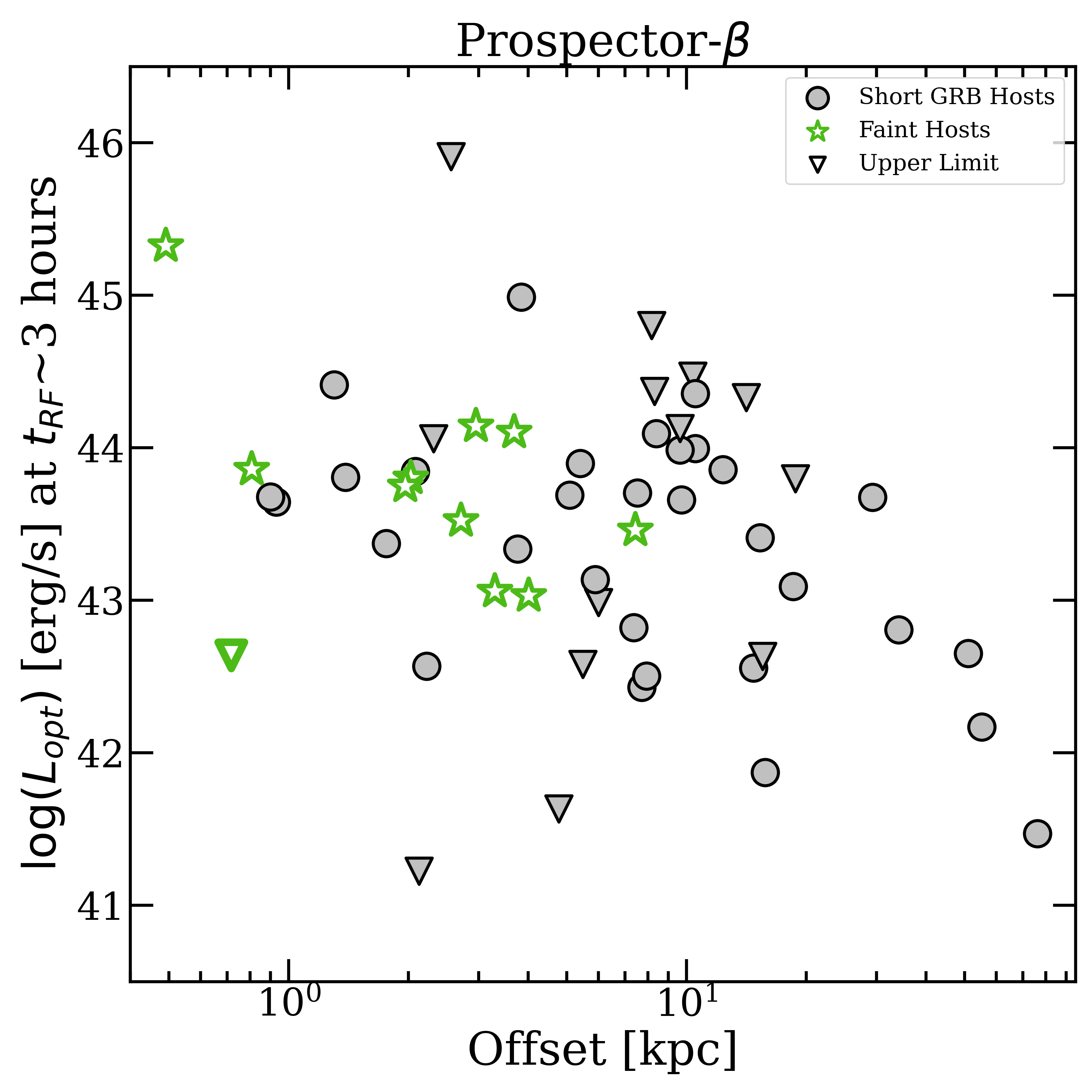}
\includegraphics[width=0.49\textwidth]{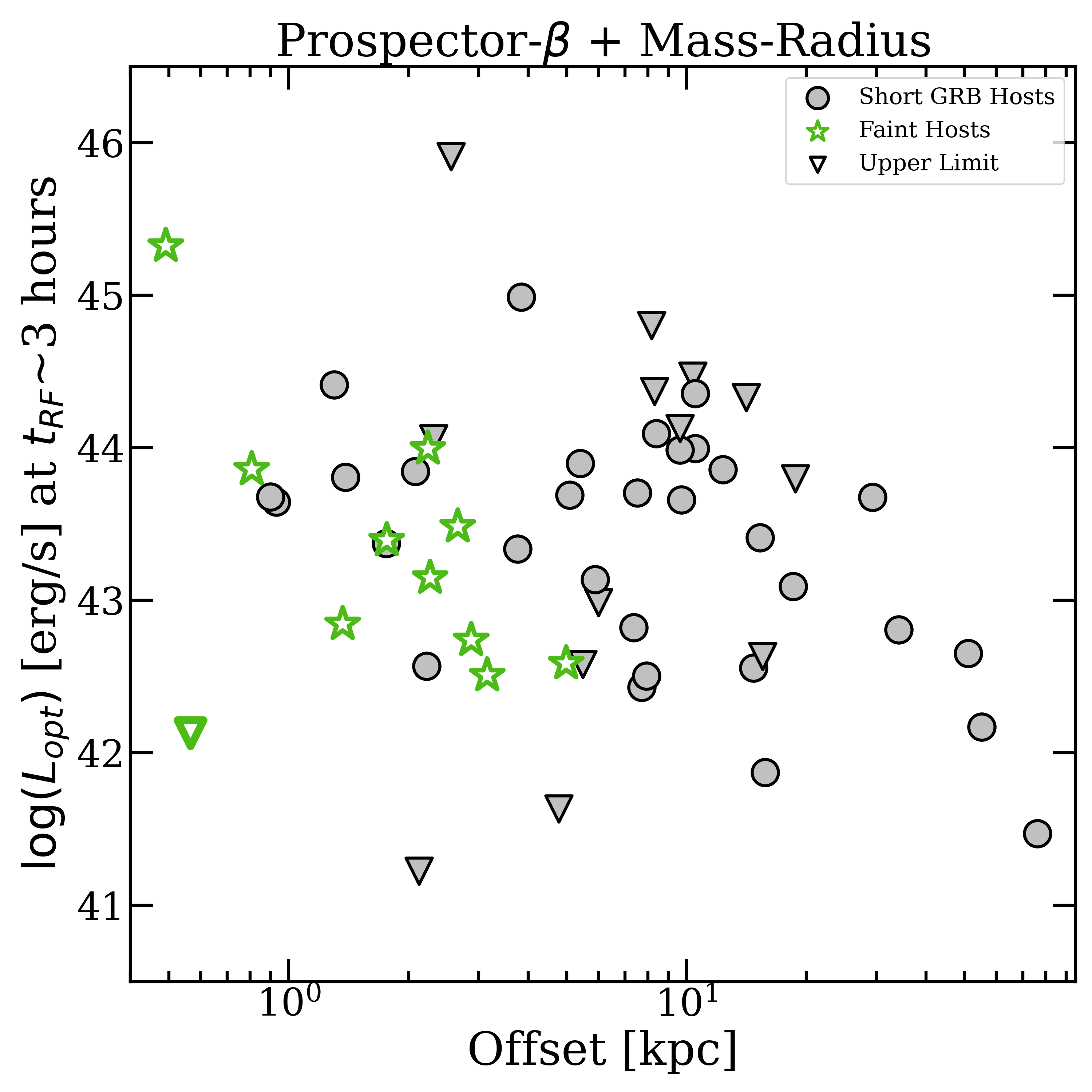}
\vspace{-0.1in}
\caption{\textit{Left:} The optical afterglow luminosities of the faint host short GRB sample (green stars) inferred from the \pbeta\ redshifts in comparison the full short GRB sample (grey circles) versus the observed physical projected offsets in kpc. Upper limits on luminosity are denoted with downward triangles. \textit{Right:} The same, but with the faint host short GRB optical afterglow luminosities determined from the \pbeta\ ($M_*$-$r_e$) redshifts. We find that the faint host short GRBs have smaller physical offsets and similar afterglow luminosities in comparison to the rest of the short GRB sample. We also see an apaprent trend of higher optical afterglow luminosities correlating to smaller offsets with this sample.}
\label{fig:afterglows}
\end{figure*}

In Figures \ref{fig:cdf_AG}-\ref{fig:afterglows}, we show optical afterglow luminosities and projected physical offsets (in kpc) of the faint host short GRBs in comparison to those of the rest of the short GRB sample from \citet{BRIGHT-I} and \citet{BRIGHT-II} at a common rest-frame time of $t_{\textrm{RF}}\sim 3$~hours. We determine luminosities at the common rest-frame time by: (i) fitting the observed afterglow data to a declining power-law model ($F_{\nu} \propto t^{\alpha}$; where $\alpha$ is typically a negative number) when there are multiple detections in a single optical filter (prioritizing $r$-band) and (ii) fitting the observed afterglow data to a $F_{\nu} \propto t^{-1}$ power-law decline when there is only one detection (e.g., $\alpha = -1$). If a short GRB has no detected optical afterglow, we include only the deepest luminosity upper limit that was $\pm 2.5$~hours from the common rest-frame time. All optical afterglow data for the faint host short GRBs are from \citet{ltf+06, ber07, pdc+08, pmg+09, gcn15224, gcn15226, gcn15312, tlt+14, kgv+17, jlw+18, rfk+2021, laskar_2022}, and all other optical afterglow data is from \citet{rfk+2021} and references therein.

We show cumulative distributions of optical afterglow luminosities for the 10 bursts from the faint host sample with detected optical afterglows (corrected to $t_{\textrm{RF}}\sim 3$~hours), along with the rest of the short GRB population (Figure~\ref{fig:cdf_AG}). We find that the faint host short GRBs have a median and population 68\% confidence interval of $\log(L_{\textrm{opt}}) \approx 43.27 \pm 0.81 $~erg~s$^{-1}$ with the \pbeta\ ($M_*$-$r_e$) redshifts\footnote{Here and onward, we use the known spectroscopic redshifts for GRBs\,090426A and 131004A in our calculations.} and a slightly higher $\log(L_{\textrm{opt}}) \approx 43.53 \pm 0.79$~erg~s$^{-1}$ from the \pbeta\ redshifts. In either case, they are similar to the optical afterglow luminosities for the rest of the population, which has a median and 68\% confidence interval of $\log(L_{\textrm{opt}}) \approx 43.65 \pm 0.79 $~erg~s$^{-1}$. To test if these differences are statistically significant, we perform an Anderson-Darling test between the distributions of detected short GRB optical afterglow luminosities. We find that $P_{AD} = 0.25$ when using the \pbeta\ ($M_*$-$r_e$) relation redshifts, and $P_{AD} = 0.16$ when using the \pbeta\ redshifts. The resulting probabilities of $P_{AD} > 0.05$ show that we cannot reject the null hypothesis that the afterglow luminosities are derived from the same underlying distribution. Thus, we find that the faint host short GRBs have statistically similar optical afterglows to the rest of the short GRB population.

However, we note that in this comparison, we are neglecting the majority of the short GRB population that only has upper limits on optical afterglow emission, and thus the gray distribution represents an upper limit. Therefore, it is possible that the optical afterglows of the faint host population are intrinsically brighter than those from more massive hosts.

\begin{figure*}[t]
\centering
\includegraphics[width=0.3\textwidth]{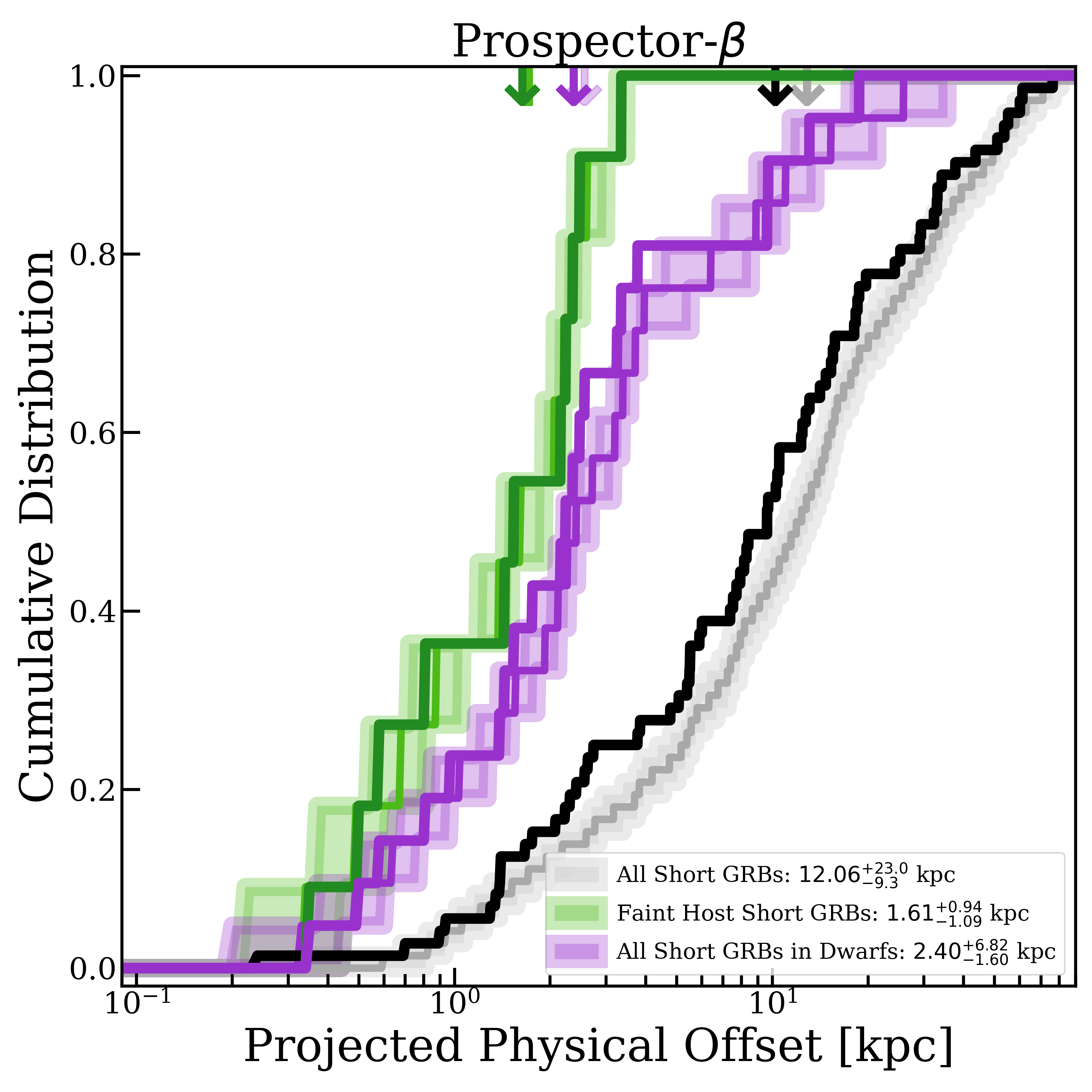}
\includegraphics[width=0.3\textwidth]{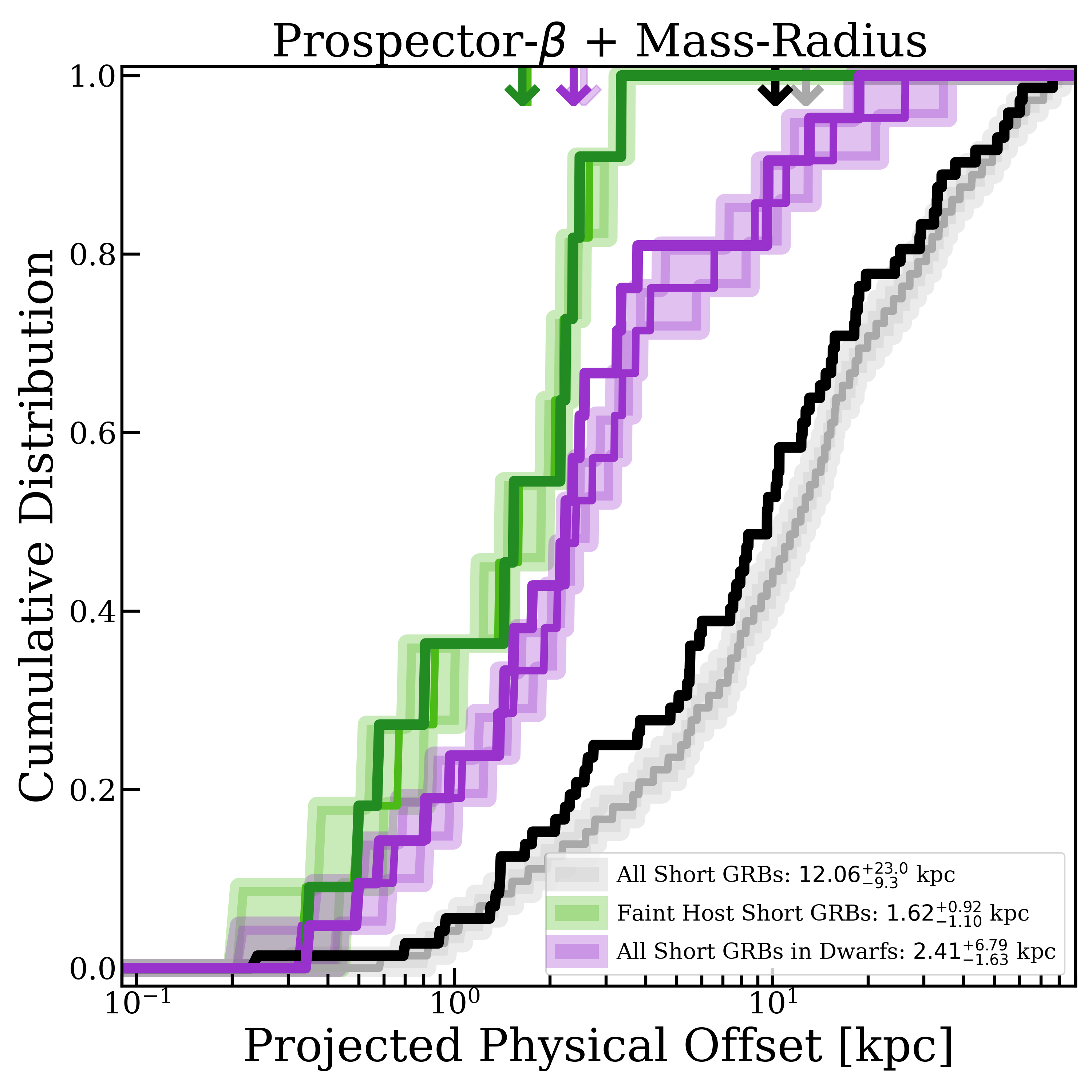}
\includegraphics[width=0.3\textwidth]{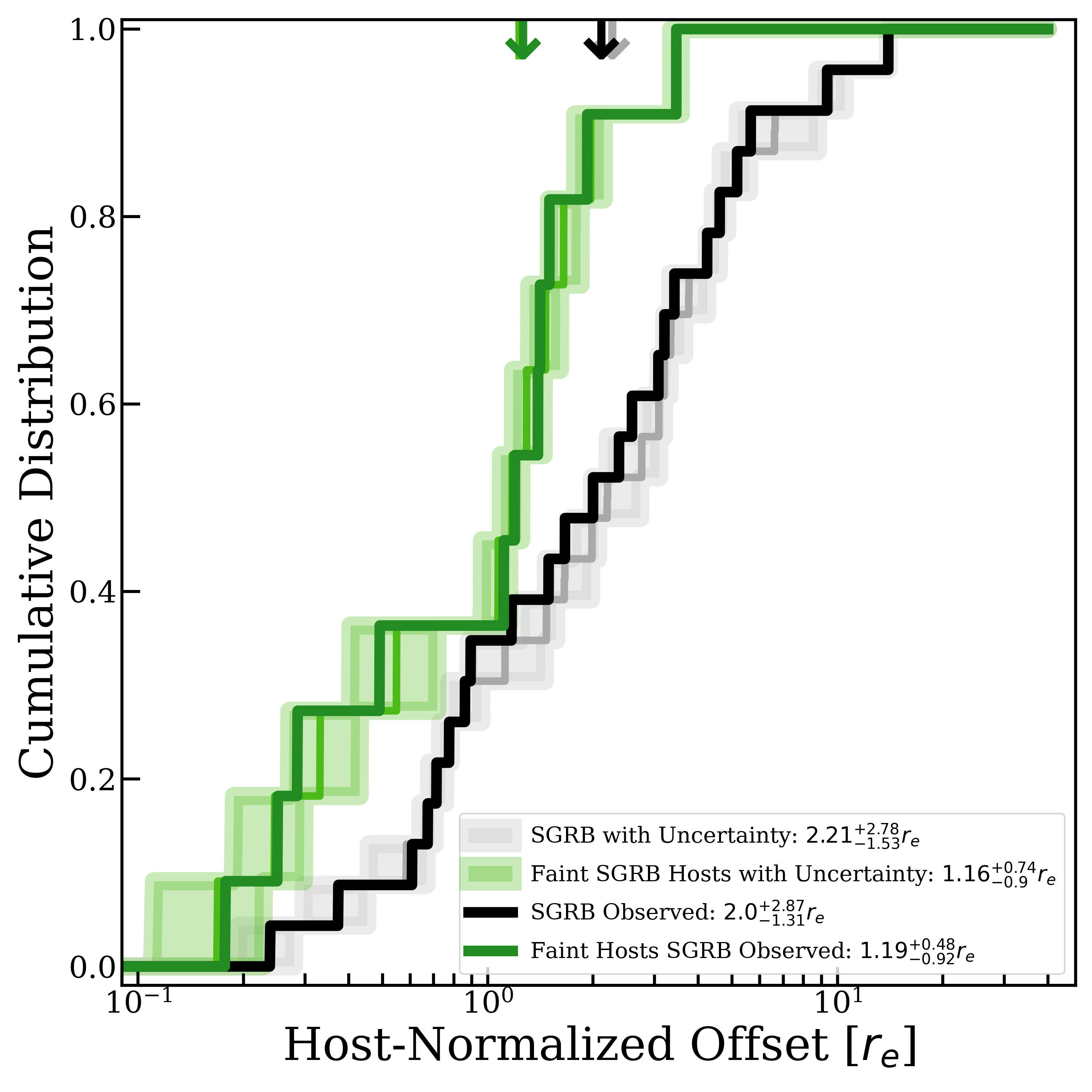}
\vspace{-0.1in}
\caption{\textit{Left:} The projected physical offset CDF and for the faint host short GRBs (observed CDF: dark green, with $1\sigma$ uncertainties CDF: light green) inferred from the \pbeta\ redshifts in comparison to that of the rest of the short GRB population  (observed CDF: black, with $1\sigma$ uncertainties CDF: grey) and all short GRBs in dwarfs (observed CDF: purple, uncertainties CDF: light purple). The legend refers to the population median and 68\% confidence interval with uncertainty. \textit{Middle:} The same, but with the faint host short GRB physical offsets inferred from the \pbeta\ and mass-radius relation redshifts. \textit{Right:} The host-normalized (galactocentric) offsets, which are independent of redshift, with the same colors as the previous plots. The arrows at the top of each plot represent the medians of each of the distributions. We find that the faint host short GRBs are both occurring at smaller physical and host-normalized offsets than the rest of the short GRB population.}
\label{fig:offsets}
\end{figure*}

We show the optical afterglow luminosities versus projected physical offsets in Figure~\ref{fig:afterglows}  (inferred from both the \pbeta\ and \pbeta\ ($M_*$-$r_e$) determined redshifts). While their afterglows appear more similar to the short GRB population, their offsets are much smaller. To more rigorously compare the offset distributions, we plot the cumulative distribution functions (CDFs) of projected physical and host-normalized offsets in Figure~\ref{fig:offsets}. We create CDFs for both the observed offsets and the observed offsets including their $1\sigma$ uncertainties, which is built from 5000 realizations on a Rice distribution (see Equation 2 in \citealt{bbf16} and Section 6.2 in \citealt{BRIGHT-I} for more details). We find for both the \pbeta\ and \pbeta\ ($M_*$-$r_e$) results, the observed projected physical offset distributions are essentially the same: the observed median lies at $1.54^{+0.86}_{-1.0}$~kpc, and when including the uncertainty, the median changes to $\approx 1.6 \pm 1.0$~kpc. The rest of the short GRB sample reside at larger projected physical offsets: $9.6^{+21.6}_{-7.5}$~kpc (observed) and $12.1^{+23.0}_{-9.3}$~kpc (with uncertainty). We note that the projected physical offsets median for the rest of the short GRB population is larger than that reported in \citet{BRIGHT-I} as we are no longer including the sample studied here in that population estimate. When we do include our sample, the median for all short GRB projected physical offsets is the same as that in \citet{BRIGHT-I}: $\approx 7.7$~kpc. We compute $P_{AD}$ between the distributions (with one test for each of the 5000 realizations on the CDF), and find that all tests result in $P_{AD} < 0.05$, demonstrating that they are statistically distinct.

We further compare the offsets for all short GRBs in a host with $M_* \leq 10^{9}\,M_{\odot}$ (a total of 21 hosts), which increases the the median of the observed offsets of short GRBs in dwarfs to $2.2^{+6.2}_{-1.4}$~kpc ($2.4^{+6.8}_{-1.6}$~kpc with uncertainties). This is plotted as the purple distribution in Figure~\ref{fig:offsets}. Anderson-Darling tests between this population and the full sample of short GRB offsets still results in $P_{AD} < 0.05$. This signifies that short GRBs identified in low-mass hosts have smaller offsets than those identified in higher-mass hosts.

When normalized by the sizes of the hosts, the same trend holds although the difference is less distinct: the faint host short GRBs have an observed median of $1.19^{+0.48}_{-0.92} r_e$ ($1.17^{+0.70}_{-0.91} r_e$ with uncertainty), and the rest of the short GRBs have an observed median of $2.0^{+2.87}_{-1.31} r_e$ ($2.21^{+2.78}_{-1.53} r_e$ with uncertainty). We find that 60\% of AD tests result in $P_{AD} < 0.05$. This suggests that the faint host short GRB hosts indeed occur closer to their host centers than the rest of the short GRB population. Given that there are higher ISM densities towards the center of galaxies, this may also explain why the majority of these short GRBs have detectable optical afterglows. We discuss implications for these results in the context of dwarf hosts in Section \ref{sec:disc}.

\subsection{Assessing Contamination from Collapsar Events}
\label{sec:contaminant}
With their low inferred stellar masses and smaller offsets, the faint host short GRB sample is reminiscent of the properties of long GRBs \citep{fls+06,kkp08, kkz+11,bbf16,llt+17,ama+20}, which typically have $\gamma$-ray duration $t_{90} \gtrsim 2$~sec and are from the collapse of massive stars (e.g., \citealt{mw99}). In particular, long GRBs are more likely to reside in dwarf and lower mass galaxies than short GRBs \citep{Svensson2010, Perley2013, Vergani2015, Wang2014, Niino2017, sys+2021, tp2021, BRIGHT-II}. Thus, a natural explanation is that the short GRBs from the faint host sample are actually collapsar events from massive stellar deaths. Here, we comment on this possibility. First, we find that the majority of the 11 GRBs with faint hosts have typical $\gamma$-ray durations spanning $0.2 \leq t_{90} \leq 1.97$~sec \citep{lsb+16}. Two GRBs in this sample (GRBs 080503 and 150424A) are classified as short GRBs with extended emission, and both have $t_{90} \geq 81$~sec.

Beyond durations, we can look at the classification criteria put forth by \citet{bnp+13} which is based on a broader range of $\gamma$-ray properties. Using the probabilities in Appendix~B of \citet{BRIGHT-I}, seven of the short GRBs in our sample have recorded $f_{NC}$ values (GRBs 060313, 090305, 090426A, 091109B, 130912A, 131004A, and 211106A), where $f_{NC}$ is the probability of not originating from a collapsar (e.g., a high probability indicates a merger origin). Of these, five have $f_{NC} \geq 0.69$, three of which have very robust, ``Gold'' host associations. The two remaining events have lower values for $f_{NC}$: GRBs\,090426A ($f_{NC} = 0.1$) and 131004A ($f_{NC} = 0.05$); these are also the two bursts with known redshift. Thus, any possible contamination from collapsar events is likely small and does not affect our overall results. 

\section{Discussion}
\label{sec:disc}

\subsection{Selection Effects}
\label{sec:disc_select}
Before discussing implications, it is useful to explore the selection effects of this sample. As a population, the 11 short GRBs studied here generally have smaller galactocentric offsets  (see Section \ref{sec:AG_offsets}) than the rest of the short GRB population. Because the probability of chance coincidence ($P_{cc}$) host association method relies on a combination of apparent magnitude and projected angular offset \citep{bkd02}, and all of the hosts in this sample are by default apparently faint, it is natural to expect that most of the short GRBs firmly associated with dwarf galaxies are at small angular offsets. Indeed, for a short GRB at the median angular offset of $\approx$1.2\arcsec\ \citep{BRIGHT-I} from a $r=27$~mag host, $P_{cc} = 0.20$ making this a fairly insecure host association. In contrast, the angular offsets of this sample are all $\lesssim 0.9\arcsec$. Thus, the discovery of a population of short GRBs with dwarf host galaxies at small offsets does not preclude a missing population at larger offsets. Indeed, one might expect that NS mergers in dwarf galaxies have larger offsets than in more massive hosts. Given the relatively shallow potential wells of dwarf galaxies, kicks at BNS formation may readily eject them from their hosts \citep{bpb+06,sra+2019}. In fact, our findings imply that we \textit{are} missing a population of dwarf hosts in current short GRB studies.

We briefly explore the current population of {\it Swift} short GRBs to search for such a population. The ten other short GRBs discovered known to have $M_*<10^9\,M_{\odot}$ dwarf hosts \citep{BRIGHT-II} have higher projected physical offsets than the sample studied here, with a population median of $3.51$~kpc, and range of 0.97--18.75~kpc \citep{BRIGHT-I}. Notably, seven of these are ``Gold" sample host associations, likely owing to the fact that these galaxies are apparently brighter than the sample studied here, and thus are easier to associate. Two of these short GRBs have offsets $>10$~kpc and could be examples of NS mergers that were ejected from their dwarf hosts. In general, the larger observed offsets for the rest of the short GRB in dwarfs sample could simply be a consequence of the galaxies having slightly larger stellar mass (median $M_* \approx 10^{8.6}\,M_\odot$), and thus larger radii, than the sample studied here. Thus, the majority of these short GRBs could still be retained within their (slightly) larger hosts but lack high-resolution data to enable host-normalized radii measurements to test this.

We further note that \citet{BRIGHT-I} accounted for six short GRBs that have inconclusive host associations (GRBs 061201, 110112A, 140516A, 150423A, 160927A, 160410A), five of which have sub-arcsecond localizations with optical afterglows. If there are underlying hosts, they all have optical magnitudes similar to the sample studied here ($\gtrsim 26$~mag), and could represent a population of dwarf hosts; however unless the angular offset is small, it would be difficult to firmly associate these with a faint host. One short GRB with an inconclusive host furthermore has a spectroscopic redshift from its afterglow: GRB 160410A at $z=1.717$ \citep{atk+2021}, with no host detected to $r>27.2$~mag. The very low inferred luminosity of $L \lesssim 10^9 L_\odot$ almost certainly puts any unseen host of GRB 160410A in the dwarf-galaxy regime. Another option is it was ejected from a more nearby, massive galaxy but no obvious candidate exists \citep{atk+2021, BRIGHT-I}. Furthermore, if the five other short GRBs are not observed to have coincident underlying galaxies, it implies that they were kicked out from a galaxy at a larger angular offset. Thus, studying the larger-scale environment surrounding these GRBs and determining if there are any dwarf galaxies in the field may better constrain the fraction of highly offset NS mergers in dwarf hosts.

\subsection{Delay Times \& Systemic Velocities}
\label{sec:disc_dt_vel}
Finding a population of small-offset short GRBs, assumed to be from NS mergers, in dwarf galaxies has unique implications for their post-supernova systemic velocities and delay times. As we mentioned in Section \ref{sec:disc_select}, we expect that dwarf galaxies in general will have very small escape velocities. Therefore, any NS merger that has been retained within a centralized location likely had a small supernova natal kick \citep{bsp99, pb02, zrd09, krz+10, brf14, wfs+18,az2019, zkn+2019}. We note, however, that this is not necessarily a straightforward comparison, as the post-supernova systemic velocity (relative to the local standard of rest at the location of the BNS-forming supernova) depends on the interplay of supernova natal kick, mass lost in the supernova, and pre-supernova orbital separation. Moreover, the distance traveled by the post-supernova NS system, and whether it escapes from the galactic potential, depends on the direction of the post-supernova systemic velocity relative to the pre-supernova galactic motion; radial post-supernova trajectories (i.e., perpendicular to the pre-supernova motion) are not necessarily the optimal means of making a system migrate far distances from a host galaxy \citep{mandel2016}. Lastly, for systems with long delay times, galactic evolution plays an important role in the kinematic evolution of kicked systems; see e.g. \citet{gw170817progenitor,zkn+2019} for deeper discussion of these effects. 

Nevertheless, we perform a simple demonstration to showcase the interplay between systemic velocities and inspiral times of the short GRB dwarf host sample studied here. If we assume that their NS merger progenitors have radial post-supernova systemic velocities that exceed their host escape velocities {\it and} that that they traveled from their hosts' effective radii towards their observed physical offset, we can derive a maximum inspiral time. To estimate the host escape velocities, we use a Hernquist density profile \citep{hernquist1990} at the median stellar mass and the median physical size derived for each host (Table \ref{tab:results}). We calculate a median escape velocity, $v_{\rm esc}\approx 14$~km~s$^{-1}$ and population range of $5.5 \leq v_{\rm esc} \leq 80$~km~s$^{-1}$, with the range based on variations of the Hernquist potential scale parameter\footnote{Here, we neglect the contribution of the dark matter halo in this simple estimate.}. We note that systemic velocities inferred for Galactic and extragalactic NS systems tend to be higher than these escape velocities. For example, Galactic NS systems have been constrained to systemic velocities ranging over 25--240~km~s$^{-1}$ \citep{fk97, wkk2000, wwk10, tkf+17}. For the short GRB population, \citet{fb13} places systemic velocity constraints ranging $20 < v < 140$~km~s$^{-1}$, while \citet{zkn+2019} finds two highly offset ($> 34$~kpc) GRBs (GRBs\,070809 and 090515) in old, quiescent, and $> 10^{10.8} M_\odot$ hosts likely have systemic velocities of $v > 200$~km~s$^{-1}$.

Using the median escape velocities of our short GRB hosts ($14$~km~s$^{-1}$) as an estimate for the smallest velocity to escape their host (corresponding to the longest inspiral time for escaping systems to reach a particular distance), and the median of their distance traveled from the hosts' effective radii ($\approx1.1$~kpc) as their radial distance, we find that the maximum inspiral is $\approx 77$~Myr. We note that if the systemic velocities were significantly higher than their host escape velocities (e.g., more similar to the observed systemic velocities of NS systems), these systems still could have merged at small offsets due to extremely short inspiral times. Without a direct probe of the inspiral time or systemic velocity, we cannot disentangle whether these systems had systemic velocities significantly lower than their host escape velocities, and thus merged at long timescales, or if they had large systemic velocities and short timescales. Constraining the SFH's of these hosts with high quality spectroscopic observations from \textit{JWST}, will lend more conclusive inferences on the inspiral times and systemic velocities of these systems.

\subsection{Implications for Gravitational Wave Follow-Up}
\label{sec:disc_gw}
Given large localization regions of mergers with current GW detectors ($> 1000$ deg$^2$ for NS mergers in the LIGO/Virgo fourth observing run; \citealt{ALIGO_det, LIGO_O4, psc+2022}), there is a need to constrain events to only their possible host galaxies to reduce the area in which to search for electromagnetic counterparts. These methods have typically relied on ranking galaxies in a field by their $B$-band luminosity, which traces star formation, or $K$-band luminosity, which traces stellar mass \citep{white2011, bbf+17, pht+17, dalya+2018, eby+2020, kfb+2022}. For NS mergers, galaxies with larger stellar mass and less star formation are generally ranked higher as possible host galaxies. Indeed, this method proved worthwhile in the case of GW170817, where its host NGC4993 was ranked high as a possible host for the event \citep{amh+2017,cfk+17, dcll2020}. However, given that short GRBs trace both star formation and stellar mass \citep{BRIGHT-II}, we do not expect their hosts to always be the brightest and most massive in field.

The addition of more dwarf hosts in this population further justifies that ranked-based methods on $B$ or $K$-band luminosities will not always be successful, as short GRBs can occur in a wider range of environments than previously thought. Assuming that we are missing a number of dwarf hosts due to larger short GRB offsets and fainter afterglow luminosities (see Section \ref{sec:disc_select}), we also may infer that the stellar mass distribution shown in Figure \ref{fig:new_mass_cdf} shifts even further towards lower stellar mass solutions. Taken together with the fact that dwarf galaxies are the most common galaxies across all redshifts \citep{bb2017}, it is quite possible that at least a fraction of NS mergers with the LIGO/Virgo GW volume will be located to dwarf hosts. Thus, galaxy-targeted searches for counterparts of GW events should be aware of this diversity when utilizing rank-based methods. 

\subsection{$r$-Process in Dwarf Galaxies}
\label{disc:r_process}
With this work, we have shown strong observational support that at least some NS mergers can occur and be retained within dwarf galaxies. These results challenge the notion that NS mergers are expected to experience strong natal kicks and long delay times, and thus likely would become unbound from dwarf hosts \citep{bhp2016, bl2016,tkf+17}. While it is not currently clear whether the small offsets we observe for this short GRB sample are due to small natal kicks, short inspiral times, or both (see Section \ref{sec:disc_dt_vel}), we can infer that one of these factors are occurring to explain our observations. Future \textit{JWST} observations of the host galaxies studied here could be used to determine their SFH's, which would then enable inferences on the delay times and systemic velocities for these short GRB progenitors. In either case, our findings lend confidence to the possibility for NS mergers to enrich dwarf galaxies with $r$-process elements, and offers a natural explanation for those observed in Local Group dwarf galaxies \citep[e.g.,][]{jfc+2016, dkak2018, mht+2021, mrr+2021, Reggiani2021, njc+2022, Limberg2023}. Our study adds to the body of work on cosmological short GRB hosts and Local Group dwarf galaxies, from which we can infer that NS mergers are an observed source of $r$-process element production in galaxies of all masses and at all redshifts.

\section{Conclusions}
\label{sec:conclusion}

In this paper, we have used novel stellar population modeling techniques to infer the redshifts and stellar masses of 11 of the faintest short GRB host galaxies, with optical and near-IR magnitudes of $>25.5$~AB~mag. This sample was selected based on their faintness in comparison to the rest of the observed short GRB host population \citep{BRIGHT-I, otd+2022}, their lack of previous redshift and/or stellar mass estimates, and their effective radii measurements from available \textit{HST} data. We list our main conclusions below.

\begin{itemize}
    \item We develop new stellar population modeling techniques that build upon the current \pbeta\ infrastructure \citep{prospector_beta}. In particular, we implement the \citet{vdw2014} mass-radius relation to more strongly weight the probabilities of \pbeta\ stellar mass and redshift solutions that result in effective radii compatible to the observed measurement. We generally find that the stellar mass estimates from this method are robust.
    \item For the 11 faint hosts studied here, assuming they are drawn from the field galaxy population, we derive a median and 68\% population confidence interval of log($M_*/M_\odot$)=$7.82^{+1.2}_{-0.86}$ and redshift at $z = 0.54^{+0.47}_{-0.29}$ when implementing the $M_*-r_e$ relation. We find stellar mass to be a fairly robust parameter, and conclude that this is a population of dwarf galaxy hosts.
    \item In comparison to the rest of the short GRB host population, these hosts have redshifts similar to the observed median ($z \approx 0.64$), but are within the bottom $\approx14\%$ of stellar mass estimates. Combined with 10 short GRBs with brighter apparent magnitudes but low stellar masses of $M_* \lesssim 10^{9}\,M_{\odot}$, we derive a lower limit on the fraction of {\it Swift} short GRBs occurring in dwarf galaxies of $\gtrsim 13.5\%$.
    \item We find that the faint host population has similar optical afterglow luminosities to the rest of the population. However, this comparison neglects the fact that the majority of short GRBs lack optical afterglow detections. Thus, when incorporating these limits, it is possible that the faint host population has more luminous afterglows.
    \item Short GRBs in faint hosts exhibit smaller projected physical and host-normalized offsets than the rest of the short GRB population. We thus infer that the majority of the short GRBs in the faint hosts were retained within their host galaxies and could represent a population of low kick velocity progenitors or those with very short delay times. This is commensurate with their (possibly) more luminous afterglows.
    \item Given that many Galactic BNS systems have velocities $\gtrsim 100$~km~s$^{-1}$, and escape velocities in dwarf galaxies are likely much lower, its possible many NS systems in dwarfs have been kicked out. Consequently, such systems exist at larger offsets, making robust host association extremely challenging. With this logic, its likely that a fraction of dwarf hosts are missing from the current short GRB host sample.
\end{itemize}

Overall, our analysis provides the first strong observational support of a  population of short GRBs in dwarf galaxies. As there is broad consensus that most short GRBs originate from NS mergers, this indicates that NS mergers are a viable source of $r$-process enrichment in these low-mass environments. Moreover, we demonstrate that at least a fraction of NS mergers can be retained in these environments. Indeed, our sample of faint host short GRBs may be representative analogs to the $r$-process event in Reticulum II and other dwarf galaxies. Our work furthermore paves the way for stellar mass and redshift estimates of faint galaxies with very sparse photometric coverage, given our novel implementation of the \citet{vdw2014} mass-radius relation into \pbeta. 

The next natural step is to obtain spectra of the sample of low-mass hosts studied here to determine their true redshifts, metallicities and star formation histories. In particular, the host galaxy SFH can constrain NS merger delay times, from which we can infer the degree of $r$-process enrichment by NS mergers in very young, low metallicity environments. Then, we can begin to answer what fraction of $r$-process elements in the Universe are derived from NS merger channels, as opposed to other proposed mechanisms. \textit{JWST} will be instrumental in the pursuit of these answers, as it has already proven to observe high quality spectra for even the faintest, highest redshift galaxies. We finally note that continued observations, follow-up, and host galaxy associations of short GRBs are needed to expand upon this faint host population and infer the fraction of dwarf hosts that are missing in current host studies. 

\section*{Acknowledgements}
We thank Bingjie Wang for fruitful discussions on \pbeta. A.E.N. acknowledges support from the Henry Luce Foundation through a Graduate Fellowship in Physics and Astronomy. The Fong Group at Northwestern acknowledges support by the National Science Foundation under grant Nos. AST-1909358, AST-2206494, AST-2308182, and CAREER grant No. AST-2047919. W.F. gratefully acknowledges support by the David and Lucile Packard Foundation, the Alfred P. Sloan Foundation, and the Research Corporation for Science Advancement through Cottrell Scholar Award 28284. C.C. acknowledges support through the Weinberg College Baker Program in Undergraduate Research.
Support for M.Z. was provided by NASA through the NASA Hubble Fellowship grant HST-HF2-51474.001-A awarded by the Space Telescope Science Institute, which is operated by the Association of Universities for Research in Astronomy, Incorporated, under NASA contract NAS5-26555. 
A.P.J. acknowledges support by the National Science Foundation under grants AST-2206264 and AST-2307599.

This research is based on observations made with the NASA/ESA Hubble Space Telescope obtained from the Space Telescope Science Institute, which is operated by the Association of Universities for Research in Astronomy, Inc., under NASA contract NAS 5–26555.

This work is based in part on observations made with the Spitzer Space Telescope, which was operated by the Jet Propulsion Laboratory, California Institute of Technology under a contract with NASA.

This work is based on observations taken by the 3D-HST Treasury Program (GO 12177 and 12328) with the NASA/ESA HST, which is operated by the Association of Universities for Research in Astronomy, Inc., under NASA contract NAS5-26555. 

Based on observations collected at the European Organisation for Astronomical Research in the Southern Hemisphere under ESO programmes 106.21T6.015, 106.21T6.016, 106.21T6.019.

This research was supported in part through the computational resources and staff contributions provided for the Quest high performance computing facility at Northwestern University which is jointly supported by the Office of the Provost, the Office for Research, and Northwestern University Information Technology.

\vspace{5mm}
\facilities{HST (WFPC2, ACS, WFC3), Spitzer, VLT (MUSE, FORS2)}
\vspace{5mm}
\software{\texttt{Prospector} \citep{Leja_2017, jlc+2021, prospector_beta}, \texttt{Python-fsps} \citep{FSPS_2009, FSPS_2010}, \texttt{Dynesty} \citep{Dynesty}}

\bibliography{refs}

\appendix 
\restartappendixnumbering

\section{Mass-Radius Relation Implementation}
\label{app:MR}
\begin{figure*}[t]
\centering
\includegraphics[width=0.49\textwidth]{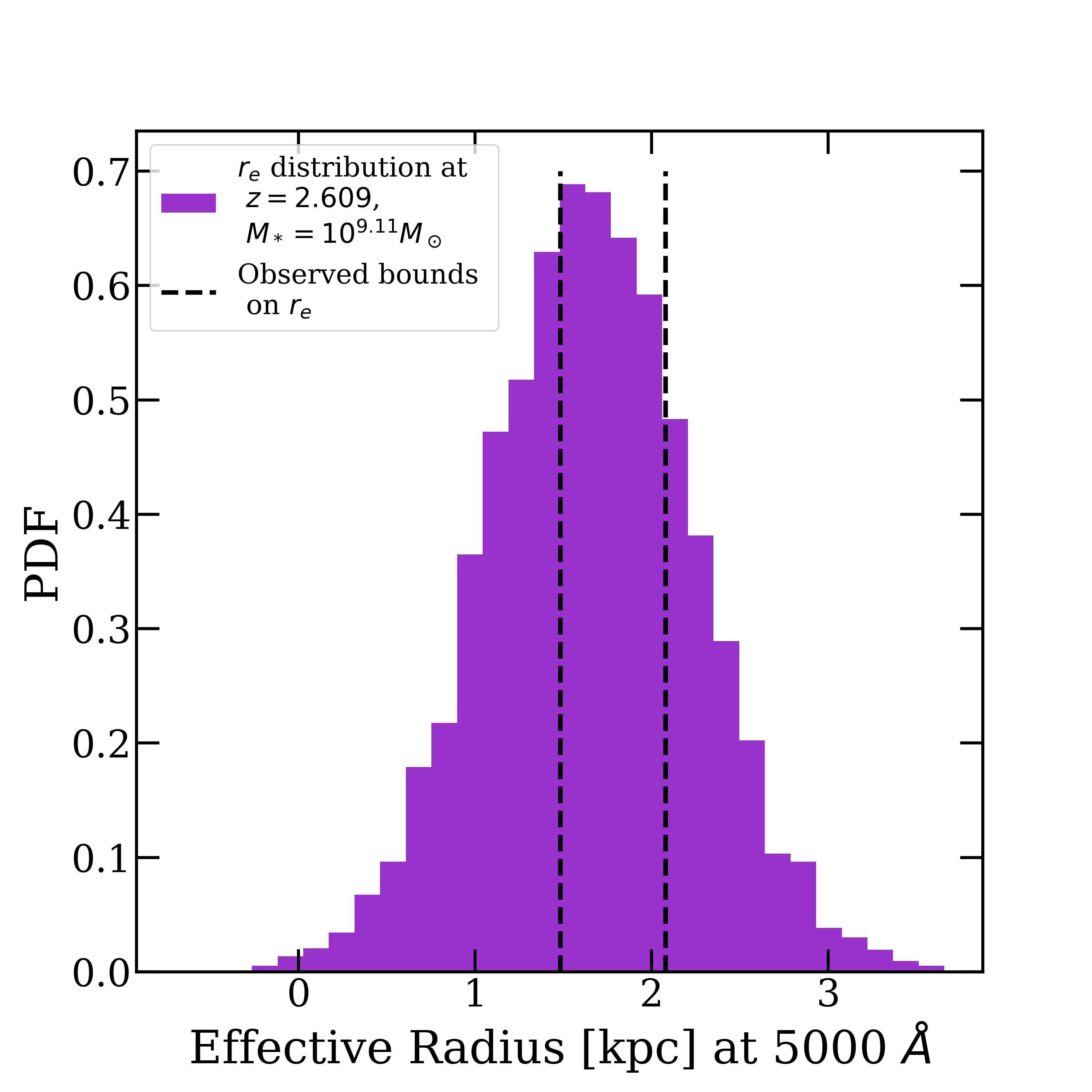}
\vspace{-0.1in}
\caption{A representation of the distribution of effective radii (purple histogram) from the \citet{vdw2014} mass-radius (Equation \ref{eqn:mass_radius}) relation at $z = 2.609$, and $M_* = 10^{9.11} M_\odot$, which are the known redshift and median stellar mass of GRB 090426A's host. We further show the bounds on the corrected effective radii at 5000 \AA\ (Equation \ref{eqn:correction}), from the observed angular effective radius of GRB 090426A's host at this redshift and stellar mass (black dashed lines). }
\label{fig:app_diag}
\end{figure*}

To correlate the \citet{vdw2014} mass-radius relation with our \texttt{Prospector}-derived results and determine which solutions are reasonable given the size of each host, we include a likelihood function of the physical $r_e$ of a host derived from a \texttt{Prospector}-sampled mass and redshift in the nested sampling fitting routine. The current nested sampling routine in \texttt{Prospector} derives the likelihood of a model SED inferred from a sample of the prior distributions, given its fit to the observed SED. Using similar logic, we include the angular $r_e$ in arcsec and the central wavelength of the filter where that measurement was taken as observed data for the host, along with its photometry. At a \texttt{Prospector}-sampled $M_F$ (which is converted to $M_*$ within the fitting) and $z$, we determine the mean physical $r_e$ and $1\sigma$ confidence interval, using: 
\begin{equation}
\label{eqn:mass_radius}
  \frac{r_e}{\textrm{kpc}} = A\times\Big(\frac{M_*}{5\times10^{10} M_\odot}\Big)^\alpha.
\end{equation}
Here, $r_e$ represents the physical $r_e$ at a rest-frame 5000~\AA, and the values for $A$, $\alpha$, and the Gaussian $1\sigma$ scatter ($\sigma\log{r_e}$) on the mass-radius relationship are given in Table 1 in \citet{vdw2014} at various redshifts over $0 < z < 3$ for late-type and early-type galaxies. We assume that the galaxies are all late-type, which is a reasonable assumption for our sample given that the hosts are likely dwarfs at moderate redshift or high redshift; in both cases, early-type galaxies are rare. Then, we convert the observed angular $r_e$ to a physical $r_e$ at $z$, and correct it to a physical $r_e$ at rest-frame 5000 \AA\ using Equation 1 in \citet{vdw2014}:
\begin{equation}
\label{eqn:correction}
\frac{\Delta \log{r_e}}{\Delta \log{\lambda}} = -0.35 + 0.12z - 0.25\log{\frac{M_*}{10^{10} M_\odot}}.
\end{equation}
Using Figure 4 in \citet{vdw2014}, we also find the $1\sigma$ scatter on this correction given the sampled stellar mass, then sample 10,000 $r_e$ from a Gaussian distribution using the mean $r_e$ from Equation \ref{eqn:correction} and the $1\sigma$ scatter. We finally determine the likelihood at each of these 10,000 observed physical $r_e$ from a Gaussian distribution given the mean and $1\sigma$ on Equation \ref{eqn:mass_radius} from the sampled $M_*$ and $z$, and determine the average likelihood. We show an example of the distribution of effective radii at a sampled $M_*$ and $z$, and the bounds determined from the mean and $1\sigma$ observed $r_e$ in Figure \ref{fig:app_diag}. The final log-likelihood at the sampled point then becomes the log-likelihood of the model SED added to the log-likelihood determined from the mass-radius relation.

\end{document}

%% file: affiliation.tex
\newcommand{\NU}{\affiliation{Center for Interdisciplinary Exploration and Research in Astrophysics (CIERA) and Department of Physics and Astronomy, Northwestern University, Evanston, IL 60208, USA}}

\newcommand{\Purdue}{\affiliation{Purdue University, 
Department of Physics and Astronomy, 525 Northwestern Avenue, West Lafayette, IN 47907, USA}}

\newcommand{\CfA}{\affiliation{Center for Astrophysics\:$|$\:Harvard \& Smithsonian, 60 Garden St. Cambridge, MA 02138, USA}}

\newcommand{\UCSC}{\affiliation{Department of Astronomy and Astrophysics, University of California, Santa Cruz, CA 95064, USA}}

\newcommand{\IS}{\affiliation{Centre for Astrophysics and Cosmology, Science Institute, University of Iceland, Dunhagi 5, 107 Reykjav\'ik, Iceland}}

\newcommand{\DAWN}{\affiliation{Cosmic Dawn Center (DAWN), Niels Bohr Institute, University of Copenhagen, Jagtvej 128, 2100 Copenhagen \O, Denmark}}

\newcommand{\PUCV}{\affiliation{Instituto de F\'isica, Pontificia Universidad Cat\'olica de Valpara\'iso, Casilla 4059, Valpara\'iso, Chile}}

\newcommand{\IPMU}{\affiliation{Kavli Institute for the Physics and Mathematics of the Universe (Kavli IPMU), 5-1-5 Kashiwanoha, Kashiwa, 277-8583, Japan}}

\newcommand{\PSU}{\affiliation{Department of Astronomy \& Astrophysics, The Pennsylvania State University, University Park, PA 16802, USA}}

\newcommand{\ICDS}{\affiliation{Institute for Computational \& Data Sciences, The Pennsylvania State University, University Park, PA, USA}}

\newcommand{\IGC}{\affiliation{Institute for Gravitation and the Cosmos, The Pennsylvania State University, University Park, PA 16802, USA}}

\newcommand{\Swin}{\affiliation{ Centre for Astrophysics and Supercomputing, Swinburne University of Technology, Hawthorn, VIC, 3122, Australia}}

\newcommand{\Curtin}{\affiliation{ International Centre for Radio Astronomy Research, Curtin University, Bentley, WA 6102, Australia}}

\newcommand{\MQ}{\affiliation{Department of Physics \& Astronomy, Macquarie University, NSW 2109, Australia}}

\newcommand{\MQAAAstro}{\affiliation{Macquarie University Research Centre for Astronomy, Astrophysics \& Astrophotonics, Sydney, NSW 2109, Australia}}

\newcommand{\CSIRO}{\affiliation{CSIRO, Space and Astronomy, PO Box 76, Epping NSW 1710 Australia}}

\newcommand{\KICP}{\affiliation{Kavli Institute for Cosmological Physics, The University of Chicago, 5640 South Ellis Avenue, Chicago, IL 60637, USA}}

\newcommand{\UChicago}{\affiliation{Department of Astronomy \& Astrophysics, University of Chicago, 5640 S Ellis Avenue, Chicago, IL 60637, USA}}

\newcommand{\UA}{\affiliation{University of Arizona, Steward Observatory, 933~N.~Cherry~Ave., Tucson, AZ 85721, USA}}

\newcommand{\EFI}{\affiliation{Enrico Fermi Institute, The University of Chicago, 933 East 56th Street, Chicago, IL 60637, USA}}

\newcommand{\mpia}{\affiliation{Max-Planck-Institut f\"{u}r Astronomie (MPIA), K\"{o}nigstuhl 17, 69117 Heidelberg, Germany}}

\newcommand{\GWU}{\affiliation{Department of Physics, The George Washington University, Washington, DC 20052, USA}}

\newcommand{\UCB}{\affiliation{Department of Astronomy, University of California, Berkeley, CA 94720-3411, USA}}

\newcommand{\RU}{\affiliation{Department of Astrophysics/IMAPP, Radboud University, PO Box 9010,
6500 GL, The Netherlands}}

\newcommand{\LJMU}{\affiliation{Astrophysics Research Institute, Liverpool John Moores University, IC2, Liverpool Science Park, 146 Brownlow Hill, Liverpool L3 5RF, UK}}

\newcommand{\LU}{\affiliation{School of Physics and Astronomy, University of Leicester, University Road, Leicester. LE1 7RH, UK}}

\newcommand{\Adler}{\affiliation{The Adler Planetarium, 1300 South DuSable Lake Shore Drive, Chicago, IL 60605, USA}}

%% file: authors.tex
\author[0000-0002-2028-9329]{Anya E. Nugent}
\NU

\author[0000-0002-7374-935X]{Wen-fai Fong}
\NU

\author[0009-0005-4757-8285]{Cristian Castrejon}
\NU

\author[0000-0001-6755-1315]{Joel Leja}
\PSU
\ICDS
\IGC

\author[0000-0002-0147-0835]{Michael Zevin}
\Adler
\NU

\author[0000-0002-4863-8842]{Alexander P. Ji}
\UChicago
\KICP

%% file: grb_obs.tex
\begin{deluxetable*}{l|cccccccc}
\tabletypesize{\footnotesize}
\tablecolumns{7}
\tablewidth{0pc}
\tablecaption{Host Galaxy Observations
\label{tab:obs}}
\tablehead{
\colhead{GRB} &
\colhead{R.A.} &
\colhead{Decl.} &
\colhead{Sample} &
\colhead{Redshift or Limit} &
\colhead{Filter} &
\colhead{AB Mag} &
\colhead{$r_e$ ($''$)} &
\colhead{References}}
\startdata
060121 & \ra{09}{09}{52.026} & \dec{+45}{39}{45.538} &  Gold & $< 4.1$ & F606W & 27.48$\pm$ 0.32 & 0.36 & 1, 2 \\
060313 & \ra{04}{26}{28.402} & \dec{-10}{50}{39.901} &  Gold & $< 1.1$ & F475W & 26.89 $\pm$ 0.20 & 0.10 & 1, 2 \\
 &  &  &  & & F775W & 26.31 $\pm$ 0.18 & & 1, 2 \\
070707 & \ra{17}{50}{58.555} & \dec{-68}{55}{27.6} &  Gold & $< 3.6$ & F606W & 26.86 $\pm$ 0.12  & 0.36 & 3 \\
&  &  &  & & F160W & 26.04 $\pm$ 0.24  & & 3 \\
080503 & \ra{19}{06}{28.901} & \dec{+68}{47}{34.78} & Silver & $< 4.2$ & F606W & 27.15 $\pm$ 0.20 & 0.26  & 4 \\
 &  & &  &  &  F160W & 26.57 $\pm$ 0.06 &   & 1, 3  \\
 &  & &  & & 3.6$\mu$m & $>$23.97 & & 5 \\
 &  & &  & & 4.5$\mu$m & $>$23.55 & & 5 \\
090305A & \ra{16}{07}{07.596} & \dec{-31}{33}{22.53} &  Gold & $< 2.9$  & F160W & 25.29 $\pm$ 0.10 & 0.36 & 3 \\
090426A & \ra{12}{36}{18.047} & \dec{+32}{59}{09.46} &  Gold & 2.609  & F160W & 25.57 $\pm$ 0.07 & 0.21 & 3 \\
 &  & & &  & 3.6$\mu$m & $>$24.58 & & 5 \\
091109B & \ra{07}{30}{56.55} & \dec{-54}{05}{23.22} & Bronze & $< 4.4$ & F110W & 27.81 $\pm$ 0.24 & 0.27 & 5 \\
130912A & \ra{03}{10}{22.2} & \dec{+13}{59}{48.74} & Bronze & $< 4.1$  & F110W & 27.47 $\pm$ 0.23 & 0.34 & 5 \\
131004A & \ra{19}{44}{27.064} & \dec{-02}{57}{30.429} & Silver & 0.717 & F110W & 25.46 $\pm$ 0.09 & 0.44 & 5 \\
150424A & \ra{10}{09}{13.406} & \dec{-26}{37}{51.745} & Silver & $< 1.1$ & F125W & 26.29 $\pm$ 0.15 &  0.28 & 5 \\
 &  & & &  & F160W & 25.89 $\pm$ 0.14 & & 5 \\
 &  & &  & & 3.6$\mu$m & $>$23.35 & & 5 \\
 211106A & \ra{22}{54}{20.541} & \dec{-53}{13}{50.548} & Gold & \nod  & $V$ &  25.45 $\pm$ 0.08 & 0.20 & 6 \\
 &  & & &  & $R$ &  26.53 $\pm$ 0.23 & & 6 \\
 &  & & &  & F814W & 25.79 $\pm$ 0.07 &   & 5 \\
 &  & &  & & F110W &  25.71 $\pm$ 0.02 & & 5 \\
\enddata
\tablecomments{The localizations, sample (confidence of host association), spectroscopic redshifts (if it is known) or maximum possible redshift of the GRB, available \textit{HST} and \textit{VLT} (MUSE $V$-band and FORS2 $R$-band) detections, \textit{Spitzer} upper limits,  and the effective radii ($r_e$) of the short GRB hosts studied in this work. Magnitudes are uncorrected for Galactic extinction in the direction of the host. \\
{\bf References:} (1) This work; (2) \citealt{fbf10}; (3) \citealt{fb13}; (4) \citealt{pmg+09}; (5) \citealt{BRIGHT-I}; (6) \citealt{fbd+2023}}
\end{deluxetable*}

%% file: results.tex
\begin{deluxetable*}{cccc|ccc}
\tabletypesize{\normalsize}
\tablecolumns{5}
\tablewidth{0pc}
\tablecaption{Stellar Population Modeling Results
\label{tab:results}}
\tablehead{
\colhead{} &
\multicolumn{3}{c}{\pbeta} & \multicolumn{3}{c}{\pbeta\ + Mass-Radius}
}
\startdata
GRB & $z$ & log($M_*/M_\odot$) & $r_e$ (kpc) & $z$ & log($M_*/M_\odot$) & $r_e$ (kpc)  \\ \hline
060121 & $1.19^{+1.4}_{-0.69}$ & $6.94^{+1.44}_{-0.91}$ & $2.87^{+0.22}_{-0.63}$ & $0.43^{+0.52}_{-0.24}$ & $8.31^{+0.91}_{-0.96}$ & $2.06^{+0.82}_{-0.88}$\\ 
060313 &  $0.49^{+0.42}_{-0.24}$ & $7.42^{+0.58}_{-0.52}$ & $0.61^{+0.18}_{-0.22}$ & $0.64^{+0.28}_{-0.32}$ & $7.36^{+0.53}_{-0.54}$ & $0.7^{+0.1}_{-0.22}$ \\
070707 &  $0.73^{+1.01}_{-0.4}$ & $7.75^{+0.56}_{-0.64}$ & $2.65^{+0.39}_{-0.93}$ & $0.29^{+0.25}_{-0.11}$ & $7.23^{+0.59}_{-0.45}$ & $1.6^{+0.73}_{-0.48}$ \\ 
080503 &  $1.05^{+1.65}_{-0.75}$ & $7.73^{+0.74}_{-0.68}$ & $2.01^{+0.2}_{-0.86}$ & $0.41^{+0.4}_{-0.19}$ & $7.01^{+0.75}_{-0.4}$ & $1.43^{+0.53}_{-0.51}$\\ 
090305 &  $0.81^{+0.59}_{-0.39}$ & $8.63^{+0.6}_{-0.63}$ & $2.77^{+0.29}_{-0.73}$ & $0.58^{+0.54}_{-0.31}$ & $8.34^{+0.85}_{-0.74}$ & $2.41^{+0.59}_{-0.89}$ \\ 
090426A &  2.609 & $9.11^{+0.33}_{-0.4}$ & 1.72 & 2.609 & $9.11^{+0.3}_{-0.39}$ & 1.72\\ 
091109B &  $0.82^{+0.69}_{-0.46}$ & $7.63^{+0.92}_{-0.82}$ & $2.06^{+0.25}_{-0.68}$ & $0.48^{+0.47}_{-0.22}$ & $7.32^{+0.95}_{-0.69}$ & $1.64^{+0.5}_{-0.53}$ \\ 
130912A & $0.81^{+0.58}_{-0.42}$ & $7.82^{+0.86}_{-0.75}$ & $2.59^{+0.3}_{-0.76}$ & $0.32^{+0.52}_{-0.14}$ & $7.28^{+1.26}_{-0.71}$ & $1.61^{+1.01}_{-0.57}$\\ 
131004A & $0.717$ & $9.07^{+0.54}_{-0.53}$ & 3.23 & $0.717$ & $9.05^{+0.38}_{-0.36}$ & 3.23 \\ 
150424A &  $0.54^{+0.28}_{-0.24}$ & $7.96^{+0.47}_{-0.49}$ & $1.8^{+0.35}_{-0.54}$ & $0.52^{+0.24}_{-0.2}$ & $7.89^{+0.48}_{-0.54}$ & $1.77^{+0.33}_{-0.46}$ \\ 
211106A & $0.73^{+0.02}_{-0.01}$ & $7.88^{+0.5}_{-0.0}$ & $1.47^{+0.0}_{-0.04}$ & $0.45^{+0.32}_{-0.0}$ & $6.98^{+1.21}_{-0.0}$ & $1.16^{+0.34}_{-0.0}$ \\ 
\enddata
\tablecomments{The results in redshift, stellar mass, and effective radii ($r_e$) from stellar population modeling fits done with \pbeta\ and \pbeta\ ($M_*-r_e$) (e.g., with the \citet{vdw2014} mass-radius relation). For the $r_e$ estimates, we convert the angular values in Table \ref{tab:obs} to physical values in kpc using the redshift posterior distribution. We note that GRBs\,090426A and 131004A have known spectroscopic redshifts via their afterglows.}
\end{deluxetable*}

%% file: resampled.tex
\begin{figure*}[t]
\centering
\includegraphics[width=0.490\textwidth]{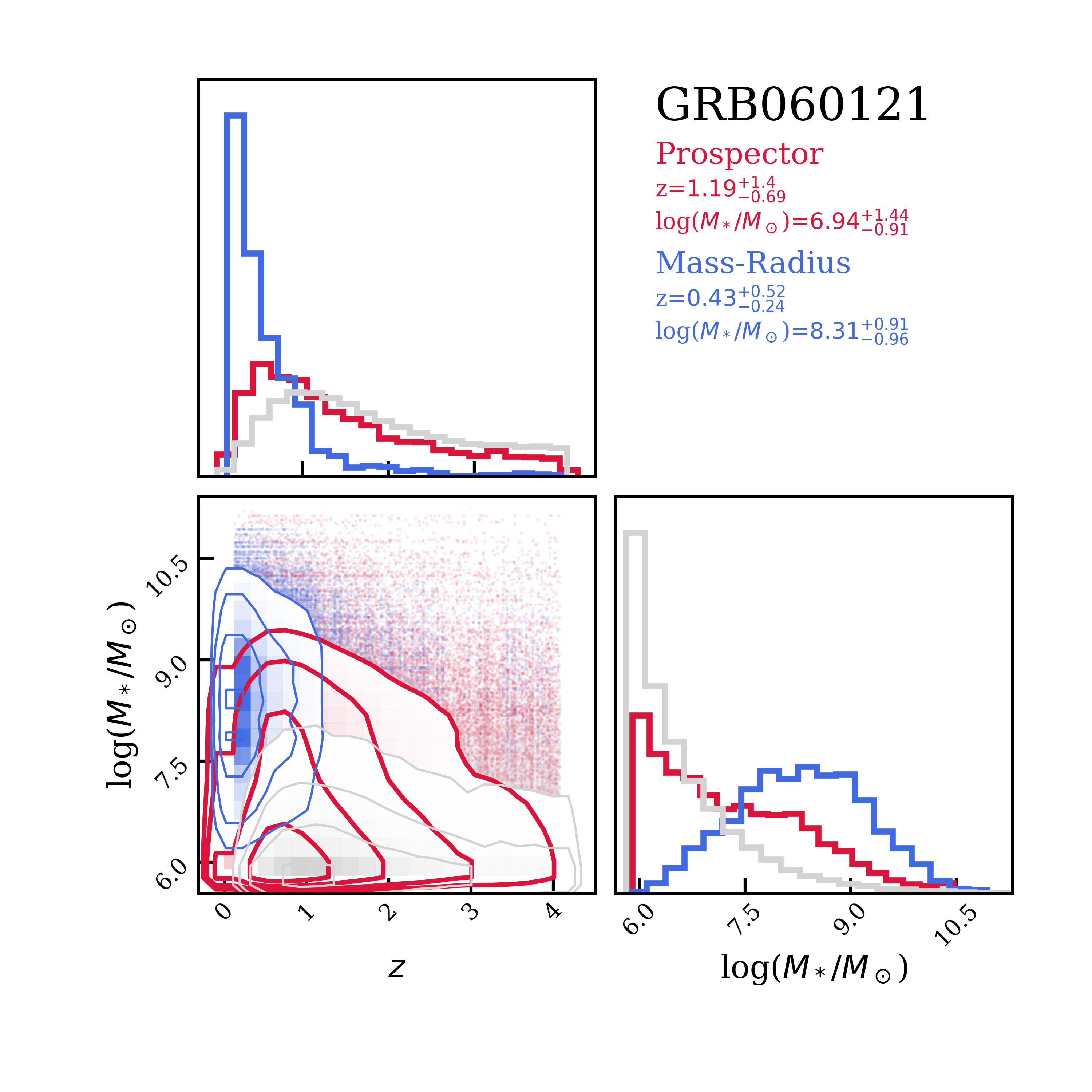}
\includegraphics[width=0.490\textwidth]{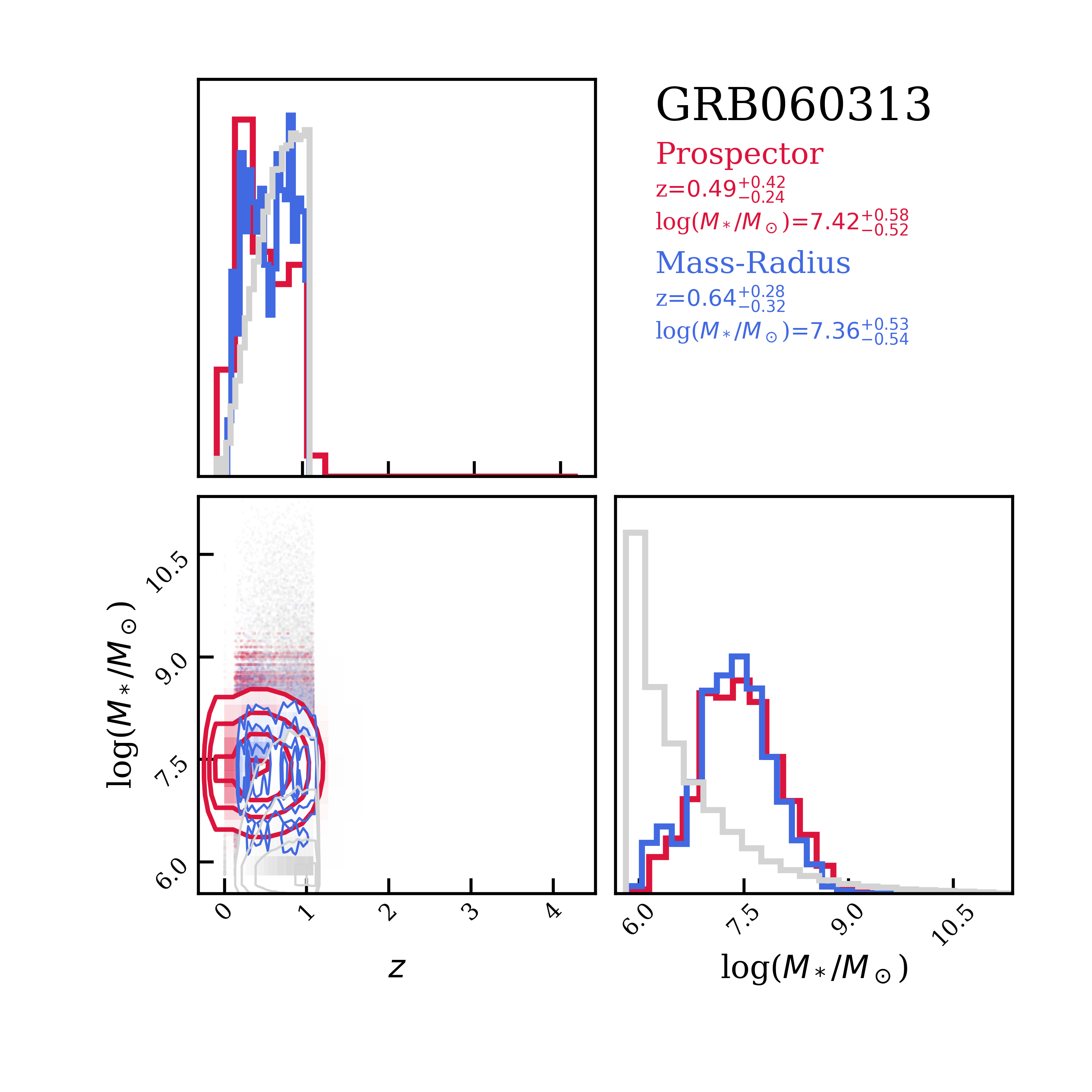}
\includegraphics[width=0.490\textwidth]{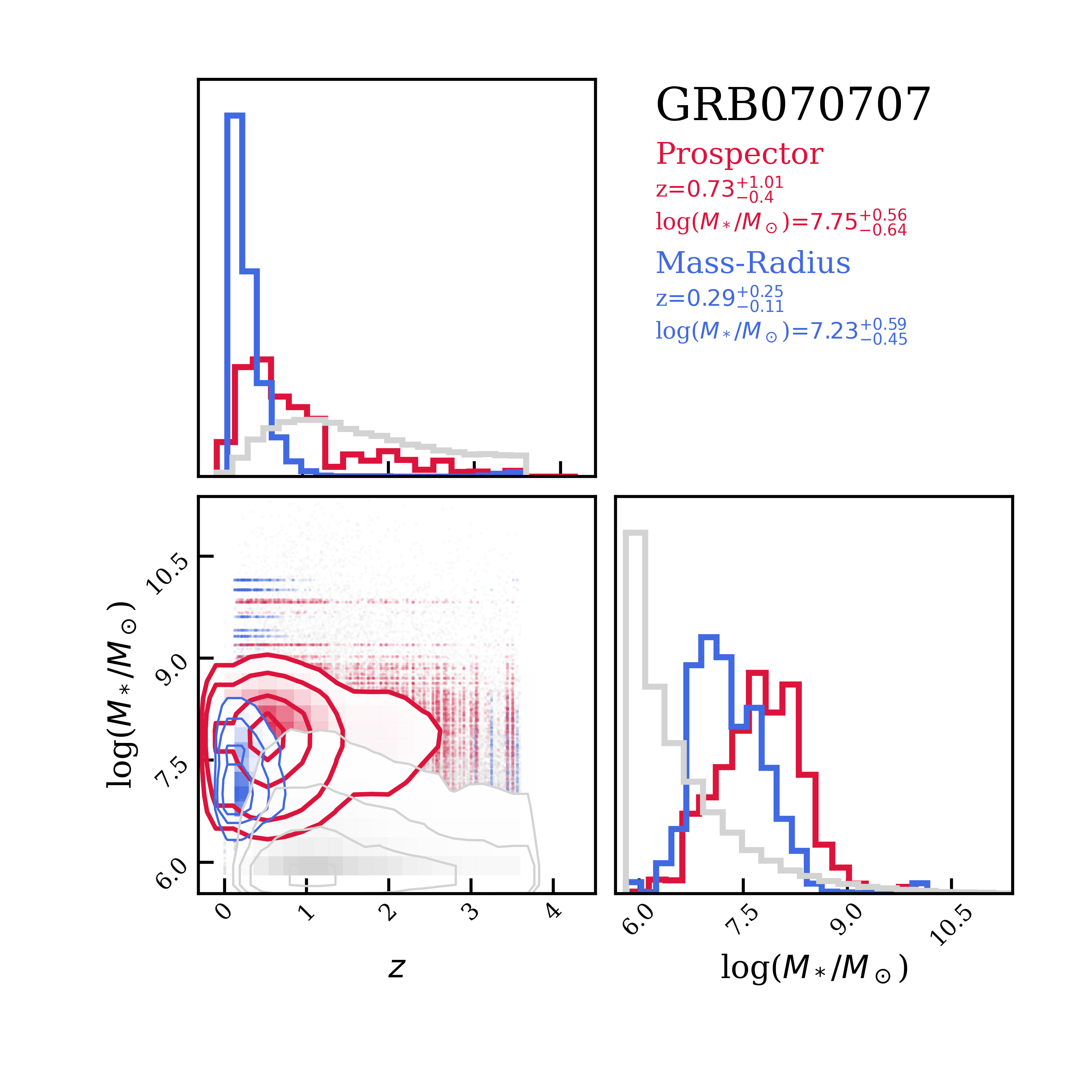}
\includegraphics[width=0.490\textwidth]{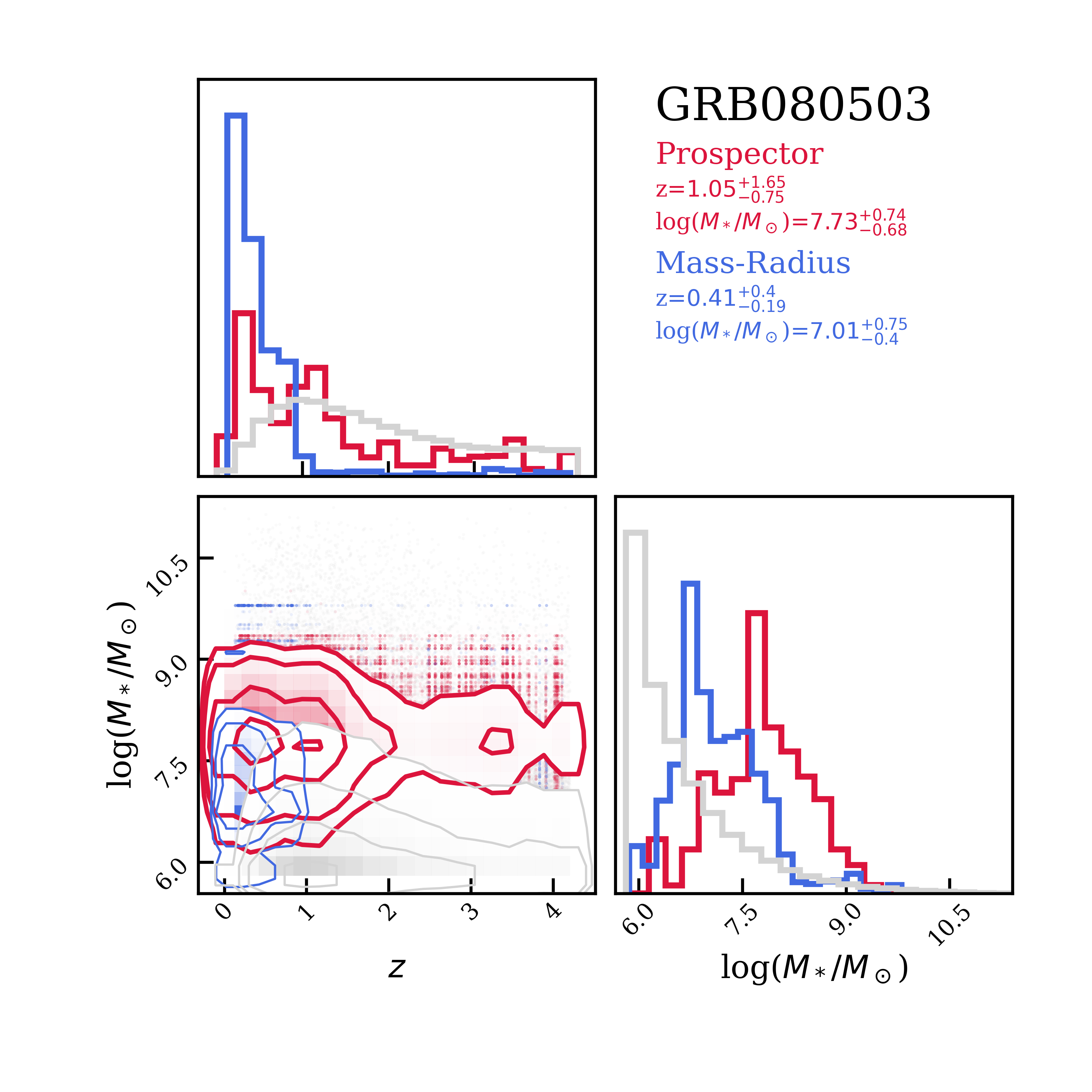}
\caption{The original \pbeta\ fits (red) compared to \pbeta\ and mass-radius fits (blue) for the sample of 9 faint short GRB hosts with no known redshifts. The prior distributions in stellar mass and redshift are shown in grey. We find that the addition of the mass-radius relation into the fits tends to lead to more constrained redshift and stellar mass estimates, with the majority of solutions leading to lower redshifts, while stellar mass estimates stay fairly similar.
\label{fig:resampled}}
\end{figure*}

\addtocounter{figure}{-1}
\renewcommand{\thefigure}{\arabic{figure} (Cont.)}

\begin{figure*}
\centering
\includegraphics[width=0.490\textwidth]{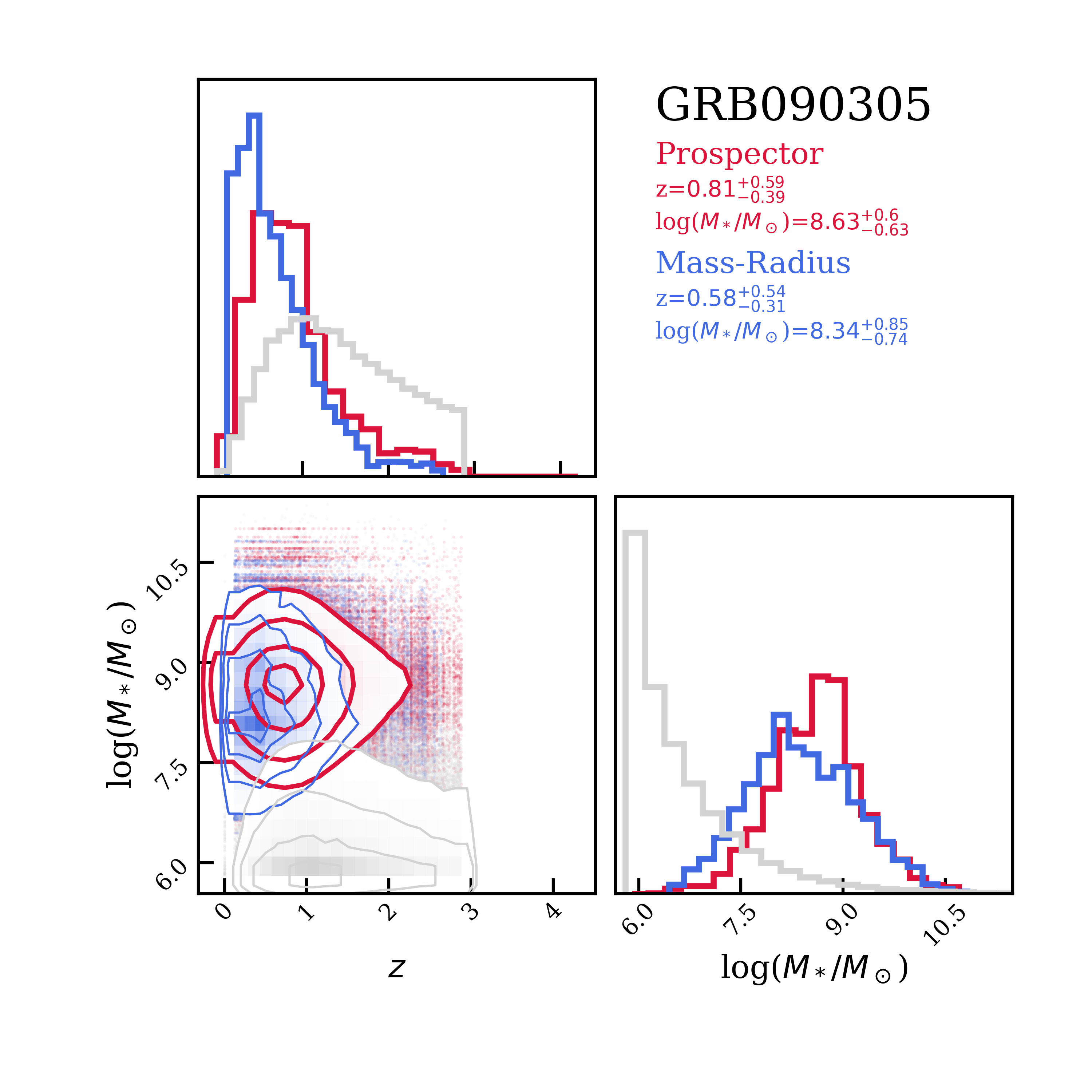}
\includegraphics[width=0.490\textwidth]{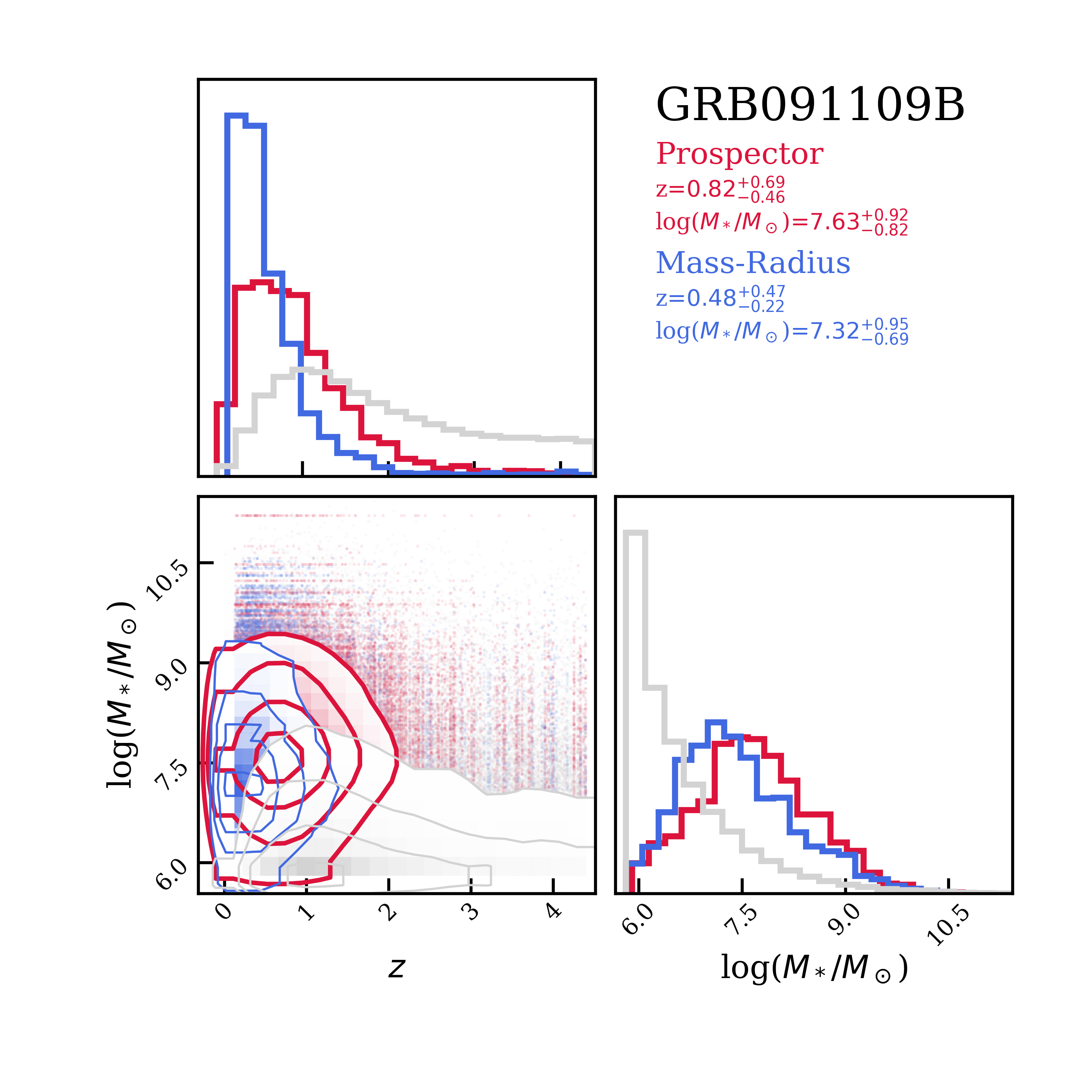}
\includegraphics[width=0.490\textwidth]{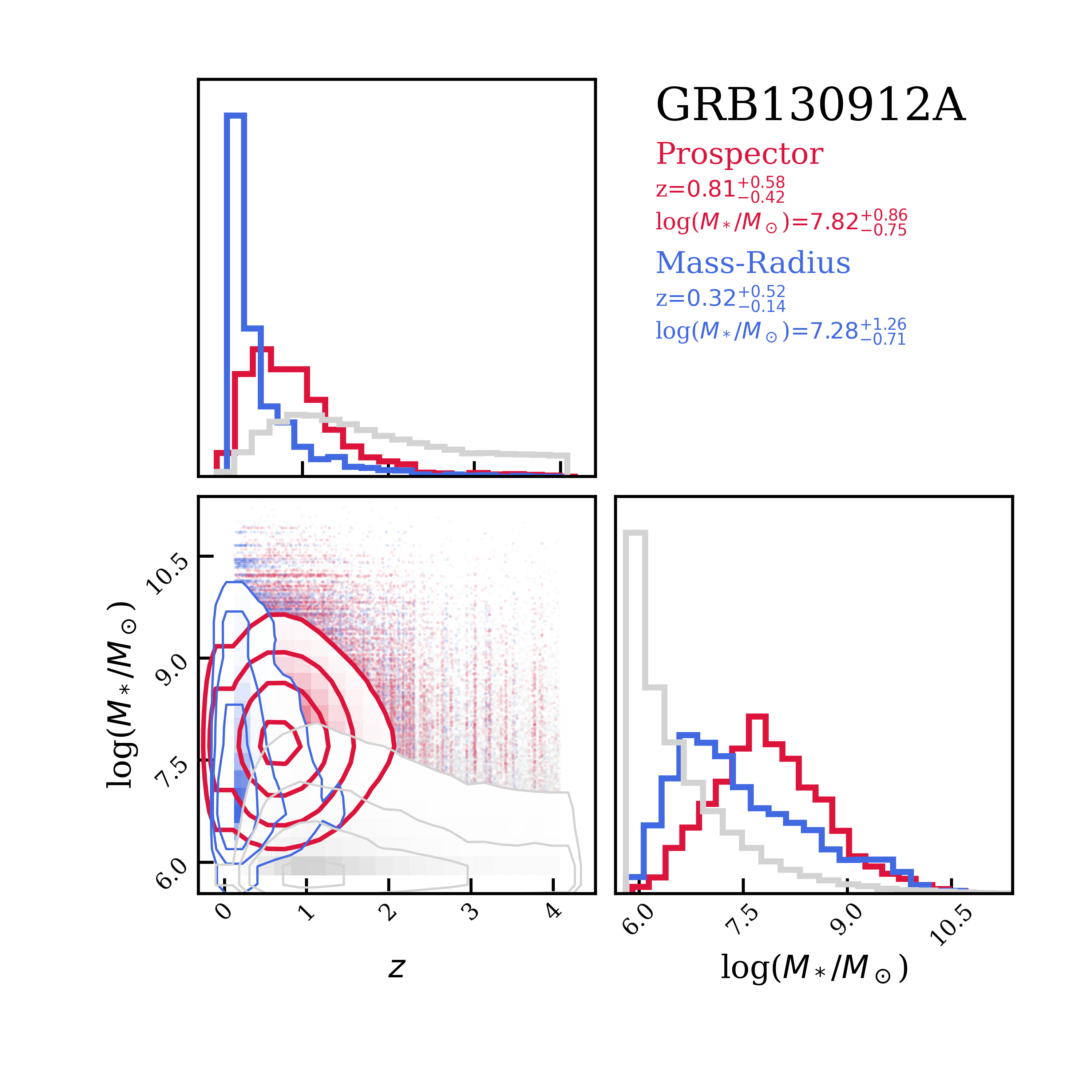}
\includegraphics[width=0.490\textwidth]{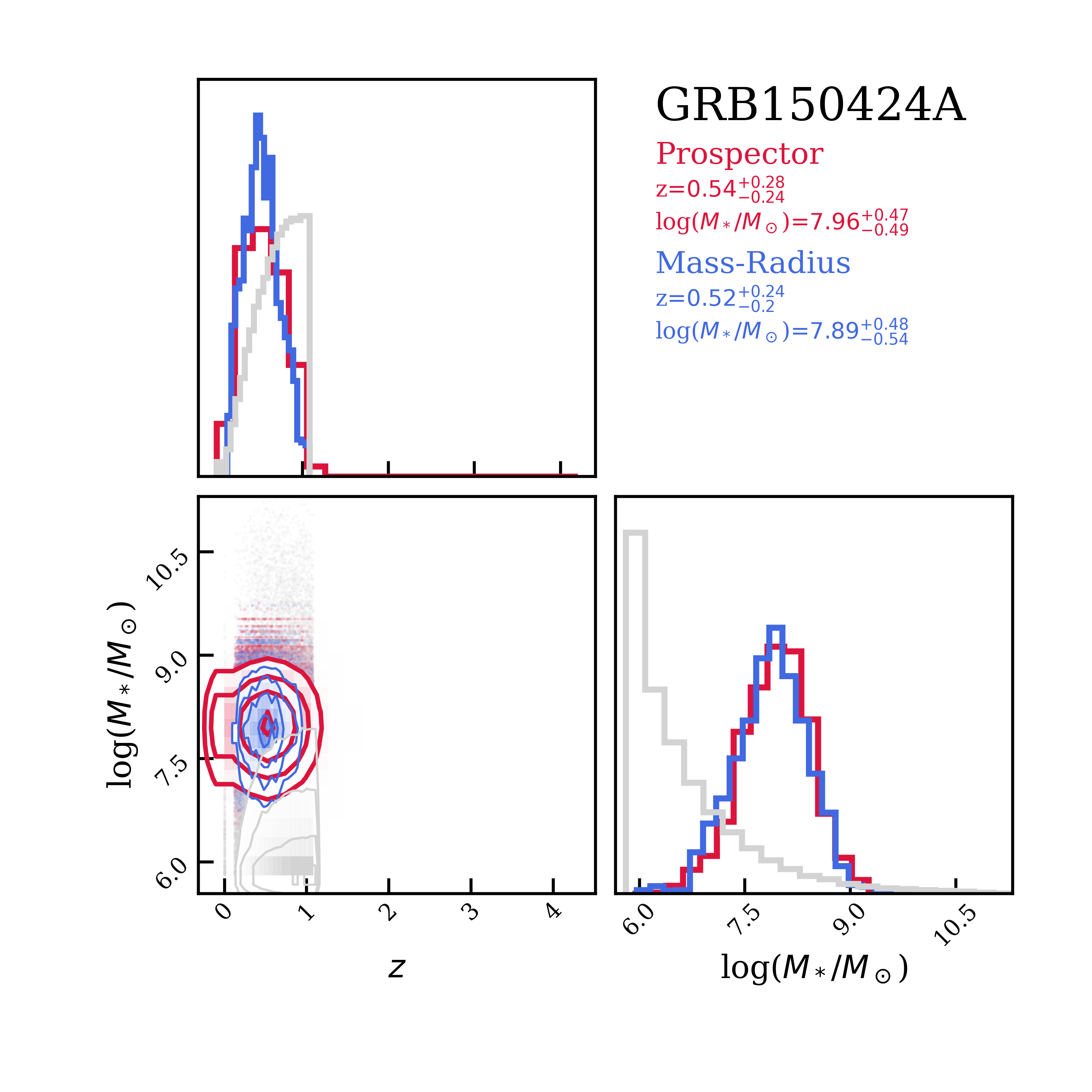}
\caption{The original \pbeta\ fits (red) compared to \pbeta\ and mass-radius fits (blue) for the sample of 9 faint short GRB hosts with no known redshifts. The prior distributions in stellar mass and redshift are shown in grey. We find that the addition of the mass-radius relation into the fits tends to lead to more constrained redshift and stellar mass estimates, with the majority of solutions leading to lower redshifts, while stellar mass estimates stay fairly similar.}
\end{figure*}

\addtocounter{figure}{-1}
\renewcommand{\thefigure}{\arabic{figure} (Cont.)}

\begin{figure*}
\centering
\includegraphics[width=0.490\textwidth]{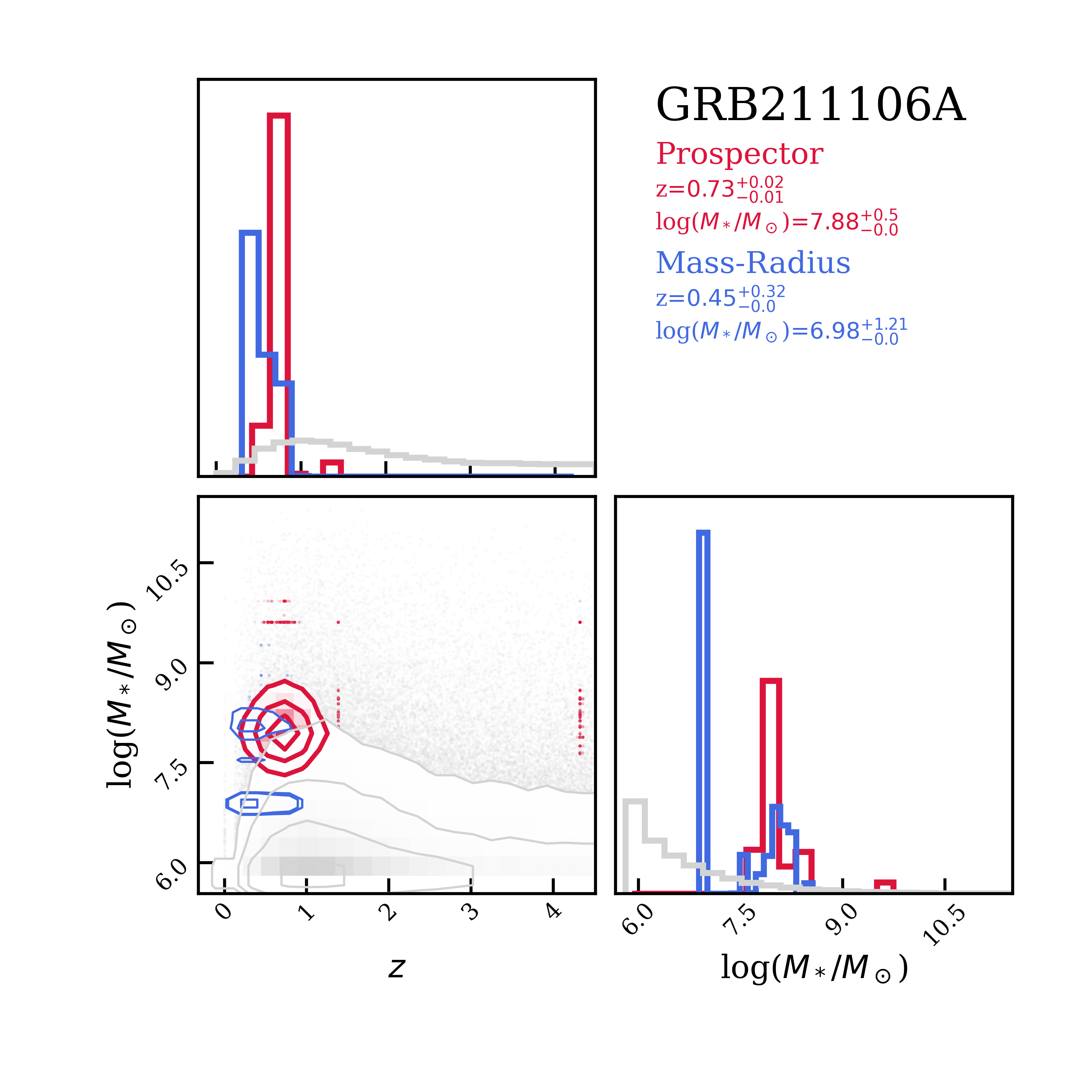}
\caption{The original \pbeta\ fits (red) compared to \pbeta\ and mass-radius fits (blue) for the sample of 9 faint short GRB hosts with no known redshifts. The prior distributions in stellar mass and redshift are shown in grey. We find that the addition of the mass-radius relation into the fits tends to lead to more constrained redshift and stellar mass estimates, with the majority of solutions leading to lower redshifts, while stellar mass estimates stay fairly similar.}
\end{figure*}

\renewcommand{\thefigure}{\arabic{figure}}